\documentclass[]{aastex631}

\usepackage{amssymb}
\usepackage[printonlyused,smaller]{acronym}

\usepackage[normalem]{ulem}

\newcommand{\mclc}{{\rm \ac{CC}}} 
\newcommand{\clc}{\ac{CC}} 
\newcommand{\image}{cutout}
\newcommand{\Image}{Cutout}
\newcommand{\animage}{a cutout}
\newcommand{\ssta}{\ac{SSTa}}
\newcommand{\sst}{\ac{SST}}

\newcommand{\syear}{2003} 
\newcommand{\eyear}{2019} 
\newcommand{\ntotal}{12,358,049} 

\newcommand{\ulmo}{{\sc ulmo}}
\newcommand{\outliers}{outlier {\image}s}
\newcommand{\mnlatent}{N_{\rm latent}} 
\newcommand{\nlatent}{$\mnlatent$}
\newcommand{\ndim}{512}
\newcommand{\nauto}{$\sim 6,000,000$}  
\newcommand{\nflow}{$\sim 22,000,000$}  

\newcommand{\fbatch}{64}  
\newcommand{\flrate}{0.00025}  

\newcommand{\mDT}{\Delta T} 
\newcommand{\DT}{$\mDT$}
\newcommand{\msT}{\sigma_T} 
\newcommand{\sT}{$\msT$}
\newcommand{\LLtenCutouts}{LL$_{10}$ cutouts}
\newcommand{\LLten}{LL$_{10}$}
\newcommand{\LLninetyCutouts}{LL$_{90}$ cutouts}
\newcommand{\LLninety}{LL$_{90}$}

\begin{document}

\title{Deep Learning of Sea Surface Temperature Patterns to Identify Ocean Extremes}

\author[0000-0002-7738-6875]{J. Xavier Prochaska}
\affiliation{Affiliate of the Department of Ocean Sciences, University of California, Santa Cruz, CA 95064, USA}
\affiliation{Department of Astronomy and Astrophysics, University of California, Santa Cruz, CA 95064, USA}
\affiliation{Kavli Institute for the Physics and Mathematics of the Universe (Kavli IPMU), 5-1-5 Kashiwanoha, Kashiwa, 277-8583, Japan}

\author{Peter C. Cornillon} 
\affiliation{
Graduate School of Oceanography, University of Rhode Island, Narragansett, RI 02882, USA; pcornillon@uri.edu}

\author{David M. Reiman}
\affiliation{
Department of Physics, University of California, Santa Cruz, CA 95064, USA; dreiman@ucsc.edu}






\begin{abstract}
We perform an \ac{OOD} analysis of 
$\sim$12,000,000 semi-independent 
128x128~pixel$^2$ \ac{SST} regions, which we define as {\image}s, 
from all nighttime granules in the
\acs{MODIS} R2019 
\acl{L2} public dataset to discover
the most complex or extreme phenomena at the ocean surface.
Our algorithm ({\ulmo})
is a \ac{PAE}, which combines two deep learning modules: 
  (1) an autoencoder, trained on $\sim$150,000 random 
  {\image}s from 2010, to represent any 
  input {\image} with 
  a \ndim-dimensional latent vector akin to a (non-linear) 
  \ac{EOF} analysis; and 
 (2) a normalizing flow, which maps the autoencoder's latent space distribution 
 onto an isotropic Gaussian manifold.  
From the latter, we calculate a \ac{LL} value for each \image\ 
and define \outliers\  to be those in 
the lowest 0.1\%\ of the distribution.
These exhibit large gradients and patterns characteristic
of a highly dynamic ocean surface, and many
are located within larger complexes
whose unique dynamics warrant future
analysis.
Without guidance, 
\ulmo\ consistently locates the outliers where
the major western boundary currents separate from the continental margin.
Buoyed by these results, we begin the process of exploring the 
fundamental patterns learned by \ulmo\, identifying several
compelling examples.
Future work may find that algorithms like \ulmo\ 
hold significant potential/promise
to learn and derive other, not-yet-identified behaviors
in the ocean from the many archives of satellite-derived \ac{SST} fields.
As important, we see no impediment to applying them 
to other large, remote-sensing datasets for ocean science
(e.g., \acs{SSH}, ocean color).

\end{abstract}



\section{Introduction}
\label{sec:intro}
\acresetall
\acused{NASA}
\acused{SSTa}


Satellite-borne sensors have for many years, been
collecting data used to estimate a broad range of meteorological, oceanographic, terrestrial and cryospheric properties. Of significance with regard to the fields associated with these properties is their global coverage and relatively high spatial (meters to tens of kilometers) and temporal (hours to tens of days) resolutions. These datasets tend to be very large, well documented and readily accessible making them ideal targets for analyses using modern machine learning techniques. Based on our knowledge of, interest in and access to global {\sst} datasets, we have chosen one of these to explore the possibilities. 
Specifically, inspired 
%
%
by the question of ``what lurks
within'' and also the desire to identify complex
and/or extreme phenomena of the upper ocean, 
we have developed an unsupervised machine learning
algorithm named {\ulmo}\footnote{{\it ULMO} is a fanciful name for our machine learning algorithm. It is based on Ulmo, the Lord of Waters and King of the Sea in J.R.R.~Tolkien's {\it Lord of the Rings}.} to analyze the
nighttime \ac{MODIS} 
\ac{L2}\footnote{`Level-2' refers to the processing level of the data, a nomenclature used extensively for satellite-derived datasets, although the precise meaning of the level of processing varies by organization. The definition used here is that promulgated by the \acf{GHRSST} - \url{https://www.ghrsst.org/ghrsst-data-services/products/}.} 
{\sst} dataset 
obtained from the \ac{NASA} spacecraft, Aqua,
spanning years \syear-\eyear.
The former (the unknown unknowns)
could reveal previously unanticipated physical processes
at or near the ocean's surface.
Such surprises are, by definition, rare and require
massive datasets and semi-automated approaches to
examine them.
The latter type (extrema) affords
an exploration of the incidence and spatial distribution
of complex phenomena across the entire ocean.
Similar `fishing' expeditions have been performed
in other fields on large imaging datasets
\cite[e.g., astronomy][]{abul21a}. However, to our
knowledge, this is the first application of machine learning for open-ended exploration of a large oceanographic dataset, although there is a rapidly growing body of literature on applying machine learning techniques to the specifics of {\sst} retrieval algorithms \cite{rs10020224}, cloud detection \cite{tc-2020-159}, eddy location \cite{moschos:hal-02470051}, prediction \cite{Ratnam:2020ta,jmse8040249,Yu:2020ur}, etc.\ and, more generally, to remote sensing \cite{ma19}. 

Previous analyses of \ac{SST} on local or 
global scales have emphasized standard statistics
(e.g., mean and RMS) and/or linear methods for pattern assessment 
(e.g., \acs{FFT} and \acs{EOF}).
While these metrics and techniques offer fundamental
measures of the \ac{SST} fields, 
they may not fully capture the complexity inherent
in the most dynamic regions of the ocean.
Motivated by advances in the analysis of natural
images in computer vision, we employ a \ac{PAE}
which utilizes a \ac{CNN} 
to learn the diversity of \ac{SST} patterns.
By design, the \ac{CNN} 
learns the features most salient to the dataset, 
with built-in methodology to examine the image
on a wide range of scales.  Further, its non-linearity
and invariance to translation offer additional advantages
over \ac{EOF} and like applications.

The \ulmo\ algorithm is a \ac{PAE}, a deep learning tool
designed for density estimation. By combining an autoencoder with a normalizing flow, the \ac{PAE} is able to approximate the likelihood function for arbitrary data while also avoiding a common downfall of flow models: their sensitivity to noisy or otherwise uninformative background features in the input \cite{nalisnick2018deep}. By first reducing our raw data (an {\sst} field) to a compact set of the most pertinent learned features via the non-linear compression of an autoencoder, the \ac{PAE} then provides an estimate of its probability by transforming the latent vector into a sample from an equal-dimension isotropic Gaussian distribution where computing the probability is trivial. We can then select the lowest probability fields as outliers or anomalous.


Our secondary goal of this manuscript, 
is to pioneer the process for like studies on
other large earth science datasets in general and oceanographic datasets in particular including 
those associated with the output of numerical models.
A similar analysis of {\sst} fields output by ocean circulation models is of particular interest as an adjunct to the work presented herein. As will become clear, we understand some of the segmentation suggested by {\ulmo} by not all of it. The method has also identified some anomalous events for which the basics physics is not clear. Assuming that the analysis of model-derived {\sst} fields yields similar results, the additional output available from the model, the vector velocity field and salinity, as well as a time series of fields, will allow for a dynamic investigation of the processes involved. 

This manuscript is organized as follows:  
Section~\ref{sec:data} describes the data analyzed
here, Section~\ref{sec:methods} details the methodology,
Section~\ref{sec:results} presents the primary results,
and Section~\ref{sec:conclusions} provides a brief
set of conclusions.
All of the software and final data products 
generated by this study are made available on-line
https://github.com/AI-for-Ocean-Science/ulmo.

\section{Data}
\label{sec:data}

With a primary goal to identify regions of the ocean exhibiting rare
yet physical phenomena, we chose to focus on the 
\ac{L2} \ac{SST} Aqua  \ac{MODIS}
dataset (https://oceancolor.gsfc.nasa.gov/data/aqua/).
 The associated five minute segments, 
each covering $\approx 2000 \times 1350$\,km of the Earth's surface 
and referred to as granules, have 
$\approx 1$\,km spatial resolution
and span the entire ocean, clouds permitting, twice daily.
For this study, we examined all nighttime granules 
from \syear-\eyear. 
The {\sst} fields, the primary element of these granules, were processed by the \ac{OBPG} at \acs{NASA}'s \acl{GSFC}, Ocean Ecology Laboratory from the \ac{MODIS} radiometric data using the R2019 retrieval algorithm \cite{Minnett2020} and were uploaded from the \ac{OBPG}'s  public server (https://oceancolor.gsfc.nasa.gov/cgi/browse.pl?sen=amod)
to the \ac{URI}.

The method developed here requires a 
set of same-sized images.
When exploring complex physical phenomena in the ocean, one is often interested in one of two spatial scales determined by the relative importance of rotation to inertia in the associated processes. The separation between these scales is generally taken to be the Rossby Radius of deformation, $R_{o}$, which, at mid-latitude is $\sim\mathcal{O}(30)$\,km. Processes with scales larger than $R_{o}$ are referred to as mesoscale processes for which the importance of rotation dominates. At smaller scales the processes are referred to as sub-mesoscale.
For this study, we chose to focus on the former
and extracted 
$128\times128$ pixel images, which we refer to as {\it cutouts},
from the \ac{MODIS} granules. Cutouts are approximately 128\,km on a side.
We are confident, supported by limited experimentation, that the
techniques described here will apply to other scales as well.

The analysis was further restricted
to data within 480~pixels of nadir.
This constraint was added to reduce the influence of pixel size on the selection process for outliers; 
the along-scan size of pixels increases away from nadir as does the rate of this increase. 
To distances of $\sim$480\,km the change in along-scan pixel size is 
less than a factor of two; at the edge of the swath the along-scan pixel size is approximately $\sim5$ times that at nadir. 

The \ac{L2}  \ac{MODIS} product  
includes a quality flag -- a measure of confidence of the retrieved \ac{SST}  -- with values from 0 (best) to 4 (value not retrieved). The primary reason for assigning a poor quality to a pixel is due to cloud contamination although there are other issues that result in a poor quality rating \cite{Kilpatrick2019}. 
A quality threshold of 2 was used for this study. 
Because the incidence, sizes, and shapes of clouds are highly variable 
(both temporally and spatially), an 
\ac{OOD} algorithm trained on 
images with some cloud contamination may become more sensitive to cloud patterns than 
unusual \ac{SST} patterns.  
Indeed, our initial experiments were stymied by clouds with the majority of 
\outliers\ showing unusual cloud patterns, 
suggesting an application of this approach to the study of clouds as well.
To mitigate this effect, we further restricted the dataset to {\image}s
with very low \clc, defined as the fraction of the
cutout image masked for clouds or other image defects.
After experimenting with model performance for various
choices of \ac{CC}, we settled on a conservative limit of
$\mclc \le 5\%$ as
a compromise between dataset size 
and our ability to further mitigate clouds (and other masked
pixels) with an
inpainting algorithm (see next section).

 

From each granule, we extracted a set of 
128x128 cutouts satisfying $\mclc \le 5\%$\ 
and distance to nadir of the central pixel $\le$480\,km.   
To well-sample the granule while limiting the number of
highly overlapping {\image}s, we drew at most
one {\image}
from a pre-defined 32x32 pixel grid on the granule.
This procedure yields $\approx 700,000$ {\image}s
per year and \ntotal~{\image}s for the full analysis.

\begin{figure}[h]
\centering
\includegraphics[width=16 cm]{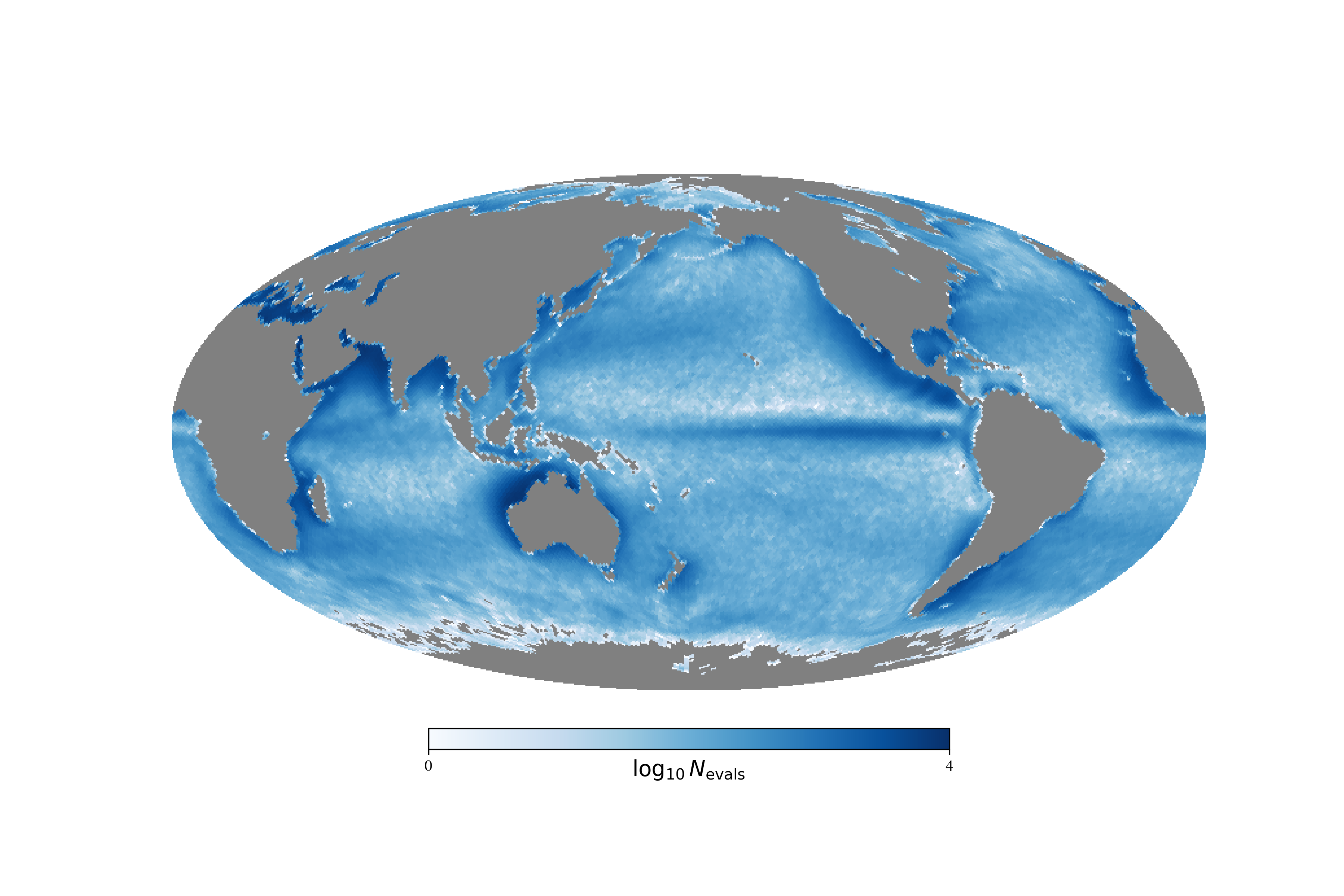}
\vskip -0.5in
\caption{Mollweide projection depicting the log$_{10}$ of the 
spatial distribution
of all {\image}s analyzed in this manuscript.  Note the higher
incidence of data closer to land driven by the lower \ac{CC} in
those areas.
}
\label{fig:spatial_full}
\end{figure}

Of course, by requiring regions largely free of clouds
($\mclc < 5\%$), we are significantly restricting the dataset
and undoubtedly biasing the regions of ocean analyzed both in time and space.
Figure~\ref{fig:spatial_full} shows the spatial
distribution of the full dataset across the ocean.
The coastal regions show the highest incidence of
multiple observations, but nearly all of the ocean 
was covered by one or more {\image}s.
Given this spatial distribution, one might naively expect
the results to be biased against coastal regions because these
were sampled at higher frequency
and comprises a greater fraction of the full
distribution.
This is mitigated, in part, by
the fact that the non-coastal regions cover a much larger area 
of the ocean but, in practice, 
we find that a majority of the 
\outliers\ are in fact located near land.

\newpage
\section{Methodology}
\label{sec:methods}

In this section, we describe the preprocessing of 
the \ac{SST}  {\image}s and the
architecture of our \ulmo\ algorithm 
designed to discover outliers within the dataset.

\subsection{Preprocessing}
\label{sec:preproc}

While modern machine learning algorithms are designed with sufficient flexibility
to learn underlying patterns, gradients, etc.\ of images \citep{inception},
standard practice is to apply initial ``preprocessing'' to each image to 
boost the performance by accentuating features of interest,
or suppressing uninteresting attributes.  
For this project, we adopted the following pre-processing steps
prior to the training and evaluation of the {\image}s.

\begin{figure}[h]
\centering
\includegraphics[width=16 cm]{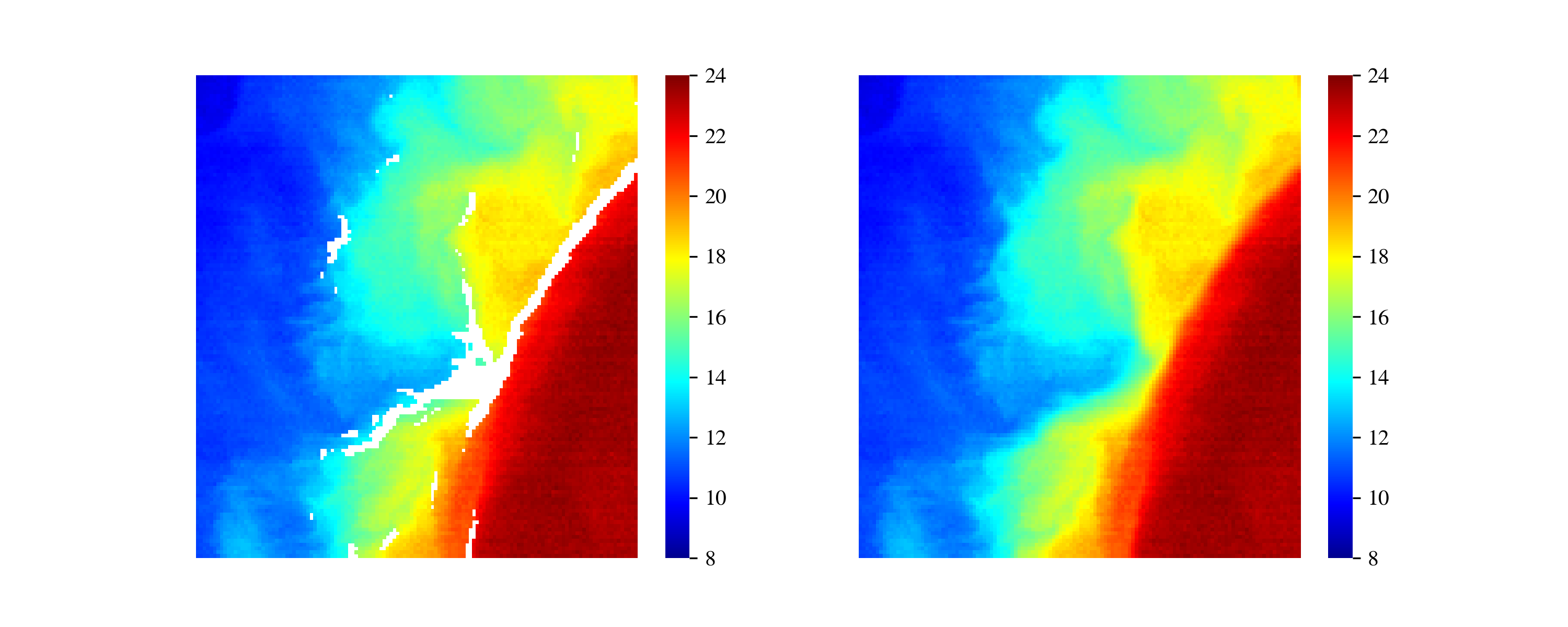} 
\caption{(left) \Image\ which shows masking (white pixels)
due primarily to sharp temperature gradients in this case, which tend to be flagged as low quality by the standard \ac{MODIS} processing algorithm.
(right) Same image but with masked pixels replaced by estimated 
values using the Navier-Stokes in-painting algorithm.
}
\label{fig:in_painting}
\end{figure}

First, we mitigated the presence of clouds.
As described in $\S$~\ref{sec:data}, this was done primarily by restricting the {\image} dataset to 
regions with $\mclc < 5\%$.
We found, however, that even a few percent cloud contamination
can significantly affect results of 
the \ac{OOD} algorithm. 
Therefore, we considered several inpainting algorithms to replace the flagged pixels
with estimated values from nearby, unmasked \ac{SST} values.  
After experimentation,
we selected the Navier-Stokes method \cite{990497}
based on its superior performance at preserving gradients within the {\image}.
Figure~\ref{fig:in_painting}
presents an example, which shows masking along a strong \ac{SST} gradient (the white pixels between the red ($\sim22^\circ$C) and yellow ($\sim19^\circ$C) regions).  
We see that the adopted algorithm
has replaced the masked data with
values that preserve the sharp, underlying gradient without
producing any obviously spurious patterns. 
Because inpainting directly modifies
the data, however, there is risk that the process will
generate {\image}s that are preferentially \ac{OOD}.
However, we have examined the set of 
\outliers\ to find that
these do not have preferentially higher \ac{CC}. 

Second we applied a 3x1 pixel median filter in the along-track direction,
which reduces the presence of striping
that is manifest in the  \ac{MODIS} \ac{L2} data product.
Third,
we resized the {\image} to 64x64 pixels using
the local mean, in anticipation of a future study on ocean
models, which have a spatial resolution of $\approx 2$\,km \cite{Qiu:2019th}.
Last, we subtracted the mean temperature from each cutout to focus the analysis on \ac{SST} differences and avoid absolute temperature being a determining characteristic. 
We refer to the mean-subtracted {\sst} values as {\ssta}.

\subsection{Architecture}
\label{sec:arch}

\ulmo\ is a probabilistic autoencoder (\ac{PAE}), a likelihood-based generative model which combines an autoencoder with a normalizing flow. In our model, a deep convolutional autoencoder reduces an input {\image} to a latent representation with \nlatent\ dimensions which is then transformed via the flow.

Flows \cite{durkan2019neural} are invertible neural networks which map samples from a data distribution to samples from a simple base distribution, solving the density estimation problem by learning to represent complicated data as samples from a familiar distribution. The likelihood of the data can then be computed using the probability of its transformed representation under the base distribution and the determinant of the Jacobian of the transformation.

Though a flow could be applied directly to image cutouts in our use case, recent research \cite{nalisnick2018deep} in the use of normalizing flows for \ac{OOD} has revealed their sensitivity to uninformative background features which skew their estimation of the likelihood. To circumvent this issue, the \ac{PAE} proposes to first reduce the input to a set of the most pertinent features via the non-linear compression of an autoencoder. The flow is then fit to the compressed representations of the image cutouts where its estimates of the likelihood are robust to the noisy or otherwise uninformative background features of the input image.

An alternative approach is the variational autoencoder (VAE) \cite{kingma2013auto} which provides a lower bound on the likelihood, though empirically we find PAEs boast faster and more stable training, and are less sensitive to the user's choice of hyperparameters.

Therefore, to summarize the advantages of our approach:
 (1) explicit parameterization of the likelihood function;
 (2) robustness of likelihood estimates to noisy and/or uninformative pixels in the input;
 and
 (3) speed and stability in training for a broad array of hyperparameter choices.
 
The key hyperparameters for the results that follow are presented in Table~\ref{tab:hyperparameters}. 
Regarding \nlatent,  we were guided by a \ac{PCA} decomposition of the
imaging dataset which showed that 512~components captured 
$> 95\%$ of the variance.
The full model
with 4096 input values per {\image}, is comprised of \nauto~parameters
for the auto-encoder and \nflow~parameters
for the normalizing flow.
It was built with PyTorch and the source
code is available on GitHub -- https://github.com/AI-for-Ocean-Science/ulmo.

\begin{table}[h]
	\centering
	\caption{Model and training hyperparameters, where the leftmost column lists the variable name, the middle column offers a brief description of the hyperparameter's function, and the rightmost column lists the value we used in our final model. Note that all autoencoder architecture hyperparameters refer to the encoder-side only (as the decoder is unused after training).}
	\label{tab:hyperparameters}
        \footnotesize
        \begin{tabular}{lll}
            \toprule
            Hyperparameter & Description & Value \\
            \hline
            \verb|n_conv_layers| & Number of convolutional layers in autoencoder & 4 \\
            \verb|kernel_size| & Size of kernel in convolutional layers & 3 \\
            \verb|stride| & Stride in convolutional layers & 2 \\
            \verb|out_channels| & Number of output channels in convolutional layers & $32 \times 2^i$ for i the layer index \\
            \verb|n_latent| & Dimension of the autoencoder latent space & 512 \\
            \verb|n_flow_layers| & Number of coupling layers in flow & 10 \\
            \verb|hidden_units| & Number of hidden units in flow layers & 256 \\
            \verb|n_blocks| & Number of residual blocks per flow layer & 2 \\
            \verb|dropout| & Dropout probability in flow layers & 0.2 \\
            \verb|use_batch_norm| & Use batch normalization in flow layers & \verb|False| \\
            \verb|conv_lr| & Autoencoder learning rate & 2.5e-03 \\
            \verb|flow_lr| & Flow learning rate & 2.5e-04 \\
            \hline
        \end{tabular}
\end{table}

\subsection{Training}
\label{sec:train}

Training of the complete model consists of two, 
independent phases: one to develop an autoencoder 
that maps input
{\image}s into \ndim-dimensional latent vectors, 
and the other to transform the
latent vectors into samples from a 
\ndim-dimensional Gaussian \ac{PDistF} to estimate their probability.
For the autoencoder, the loss function is the standard 
mean squared error reconstruction loss between all pixels in the input and output {\image}s. In practice, the model converged to a small loss in $\sim 10$~epochs of training.

The flow is trained by directly maximizing the likelihood of the autoencoder latent vectors. This equates to minimizing the Kullback-Leibler divergence between the data distribution and flow's approximate distribution. Minimizing this divergence encourages the flow to fit the data distribution and thereby produce meaningful estimates of probability.

Throughout training, we used a random subset of $\approx 20\%$
of the data from 2010
(135,680 {\image}s).  These {\image}s
were only used for training and are not evaluated in
any of the following results. 

\begin{figure}[h]
\centering
\includegraphics[width=16 cm]{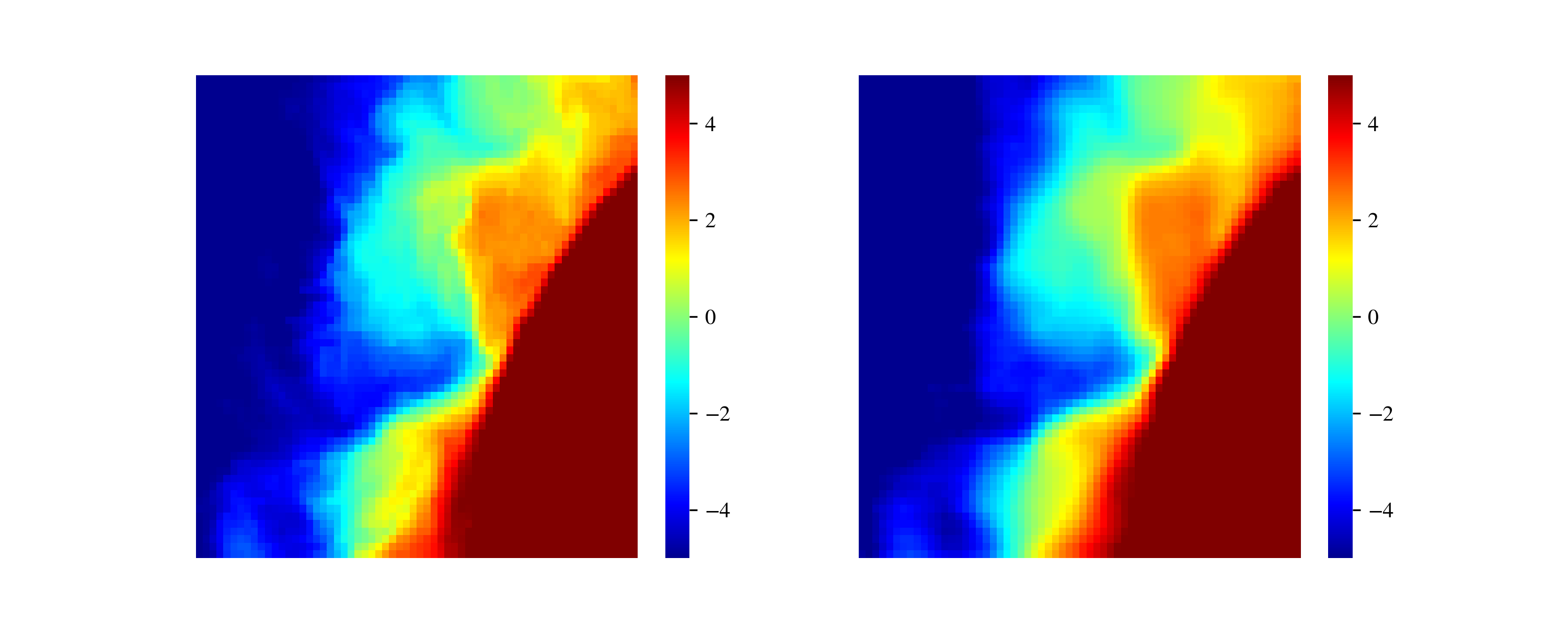} 
\vskip -0.2in
\caption{
(left) Input \image, preprocessed as described in $\S$~\ref{sec:preproc}.
(right) Image generated by the trained autoencoder, 
which passes
the \image\ through a 512~unit latent space.
As designed, the reproduction captures the gross features but not all of the small-scale detail.
}
\label{fig:auto}
\end{figure}

Figure~\ref{fig:auto}
shows an example
of a preprocessed input {\ssta} {\image} and 
the resultant reconstruction {\image} from
the autoencoder.  As designed, the output is a good
reconstruction albeit at a lower resolution
that does not capture all of the finer features due to the information bottleneck in the autoencoder's latent space
but it does capture the mesoscale structure of the field.
For the normalizing flow, we used {\animage}
batch size of \fbatch\ and a learning
rate of \flrate.  Similarly, we found
$\approx 10$~epochs were sufficient to achieve
convergence.

We performed training on the Nautilus distributed computing system
with a single GPU. In this training setup, a single epoch
for the auto-encoder requires 100\,s while a single epoch for the flow requires $\approx 900$\,s.

\section{Results and Discussion}
\label{sec:results}

In this section, we report on the main results of our analysis
with primary emphasis on outlier detection.
We also begin an exploration of the \ulmo\ model to better understand
the implications of deep learning for analyzing 
remote-sensing imaging;
these will be expanded upon in future works.

\begin{figure}[h]
\centering
\includegraphics[width=16 cm]{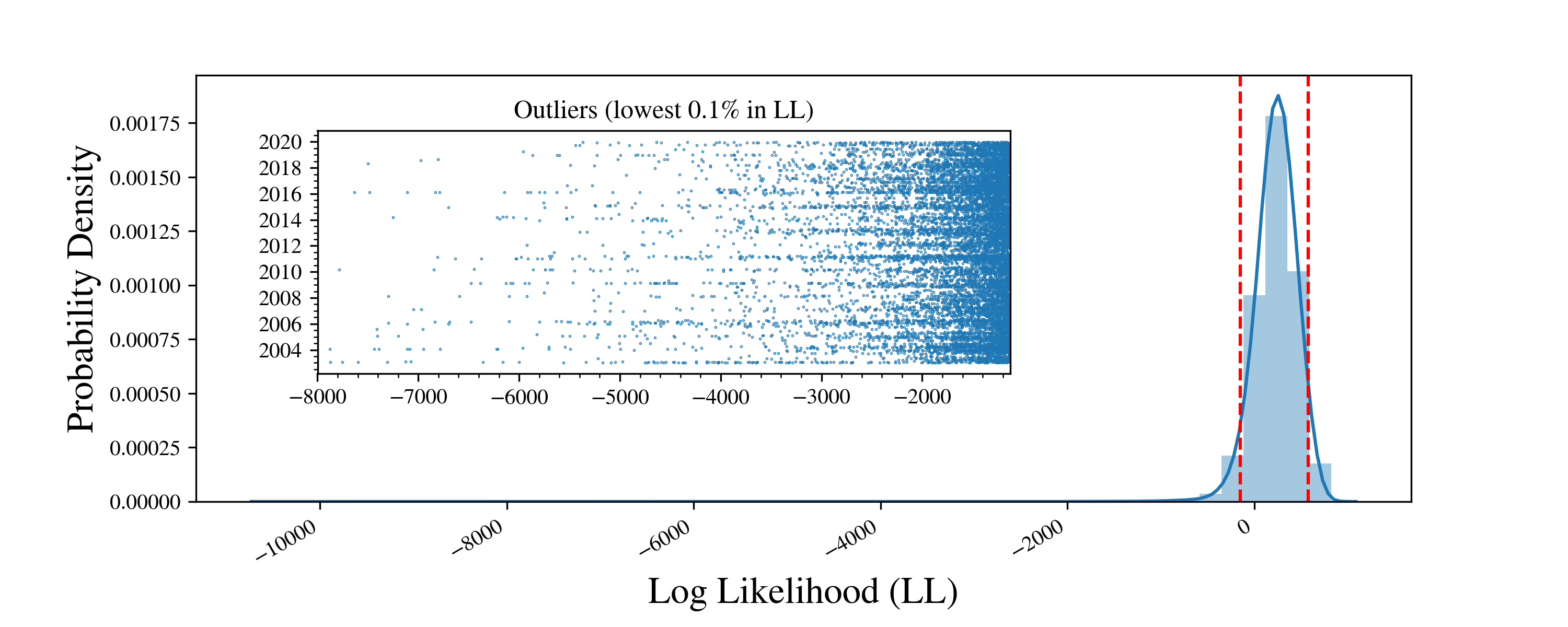}
\caption{
Distribution of \ac{LL} values for the full dataset 
(modulo the training images).  The majority are well described
by a Gaussian centered at ${\rm LL} \approx 240$,
with a tail to much lower \ac{LL} values. 
The inset shows the lowest 0.1\%\ sample of LL, 
which we define as \outliers.  
}
\label{fig:LL_ssta}
\end{figure}   

\subsection{The \outliers\ sample}
\label{sec:anomalies}

Figure~\ref{fig:LL_ssta} shows the \ac{LL}
distribution for all extracted {\image}s 
modulo the set of training {\image}s from 2010.
The distribution peaks at \ac{LL}~$\approx 240$
with a tail to very low values.
The latter is presented in 
the inset which shows the lowest 0.1\%\ 
of the distribution; these define
the \outliers\ of the full sample (or outliers for short).

\begin{figure}[h]
\centering
\includegraphics[width=12 cm]{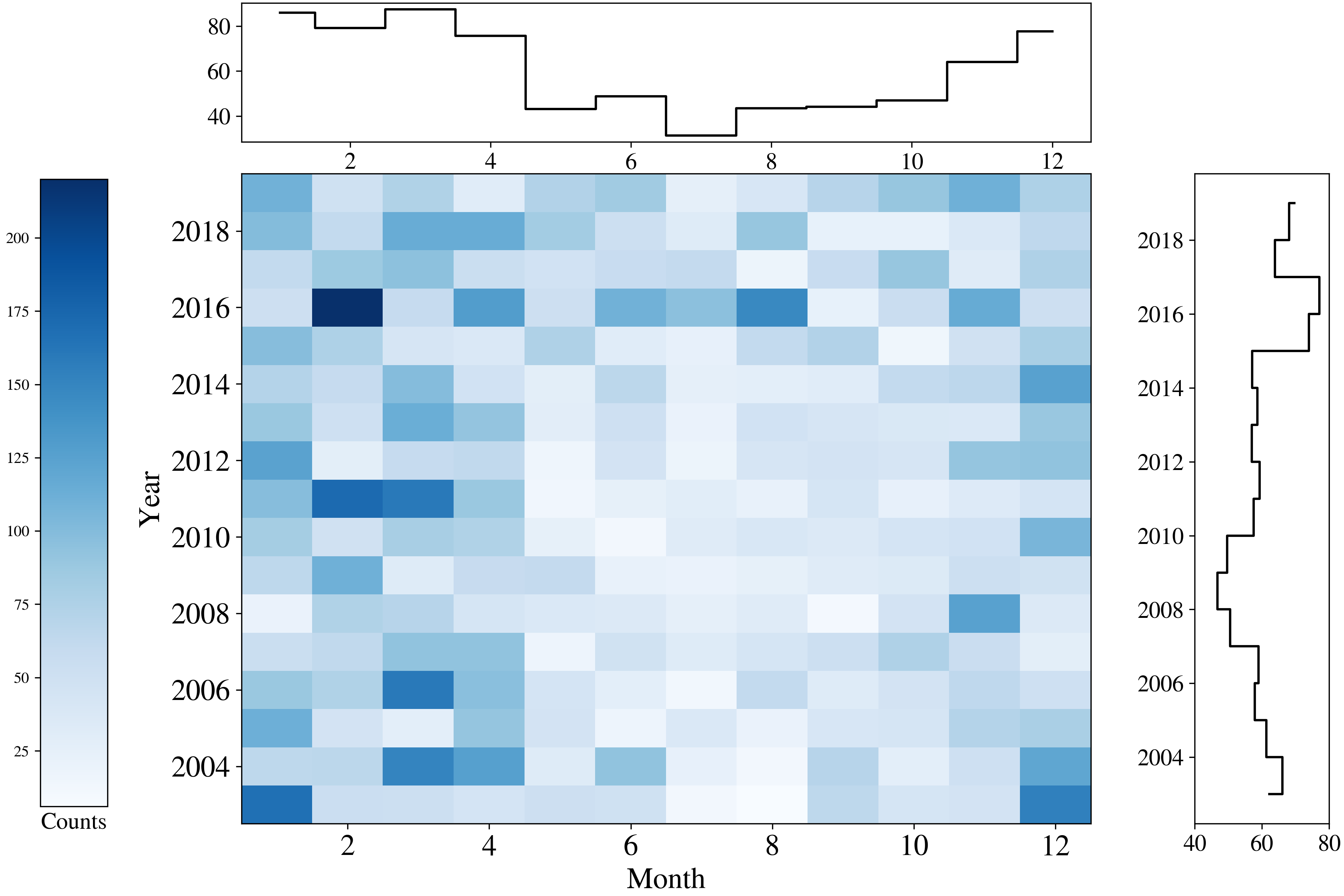}
\caption{Incidence (counts) of \outliers\ broken down 
by month and year.  The primary feature is seasonal,
i.e., a higher number of \outliers\ during boreal winter
months than summer.  There is also a weak, but possible
increase in the incidence of outliers over the past $\sim 10$~years.
}
\label{fig:temporal}
\end{figure}   

The striping apparent in the inset of Figure~\ref{fig:LL_ssta}
indicates a non-uniform,
temporal dependence in the outlier {\image}s.
Figure~\ref{fig:temporal} examines this further, 
plotting the
occurrence of outliers as a function of 
year and month.  
The only significant trend apparent is seasonal,
i.e., a higher incidence of outliers during the boreal winter. 
We speculate this is due to the predominance of northern hemisphere cutouts/outliers -- approximately 60\%/64\% of the total -- and the reduced thermal contrast of northern hemisphere surface waters in the boreal summer. As will be shown, the range of {\ssta} in a cutout is correlated with the probability of the cutout being identified as an outlier; the larger the range the more likely the cutout will be so flagged. This is especially true in the vicinity of strong currents such as western boundary currents, which separate relatively warm, poleward moving equatorial and subtropical waters from cooler water poleward of the currents. In summer months the cooler water warms substantially faster than the surface water of the current dramatically reducing the contrast between the two water bodies, often masking the dynamical nature of the field in these regions rendering them less atypical.
We also see variations during the $\sim 20$~years of the 
full dataset, including a possible increase over the past $\sim 10$~years.
These modest trends aside, \ulmo\ identifies outliers 
in all months and years of the dataset.

\begin{figure}[h]
\centering
\includegraphics[width=16 cm]{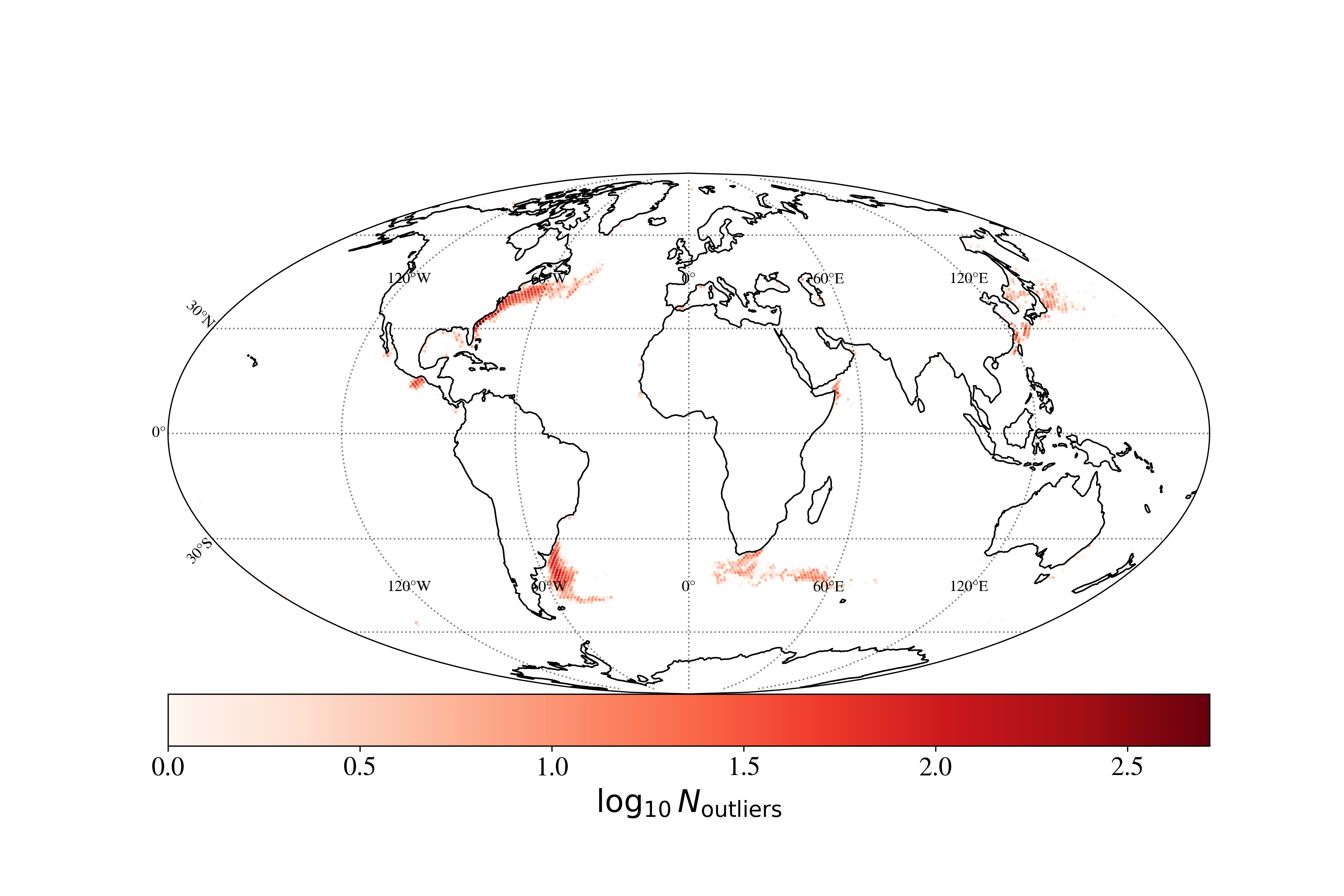}
\vskip -0.3in
\caption{Depiction of the spatial distribution
of the outliers discovered by \ulmo.
These are primarily in the well-known 
western boundary
currents off Japan, North and South America,
and South Africa.
Note that the scaling is logarithmic.
}
\label{fig:OOD_spatial}
\end{figure}   

A question that naturally arises is whether there is any 
structure to the geographic distribution of outliers.
Figure~\ref{fig:OOD_spatial}
shows the count distribution of the
outliers across the entire ocean.
Remarkably, the \ulmo\ algorithm
has rediscovered that the rarest phenomena
occur primarily in western boundary 
currents -- following the continental boundary and/or shortly after separation.
These regions of the ocean have been studied extensively because
of their highly dynamical nature.
In short, the \ulmo\ algorithm identified (or even rediscovered!)
without any predisposition 
a consistent set of dynamically important oceanographic regions.

To a lesser extent, 
one also finds outliers
in the vicinity of the connection 
between large gulfs or seas and the open ocean -- the Gulf of California, the Red Sea and the Mediterranean. 
Also of interest are the outliers in the Gulf of Tehuantepec. These result from very strong winds blowing from the Gulf of Mexico to the Pacific Ocean through the Chivela Pass, resulting in significant mixing of the near-shore waters. 

There are two ways to view the results in
Figure~\ref{fig:OOD_spatial}: 
 (1) as the contrarian, i.e., the \ulmo\ algorithm
 has simply reproduced decades-old, basic knowledge 
 in physical
 oceanography on where the most dynamical regions
 of the ocean lie; or 
 (2) as the optimist, i.e., the \ulmo\ algorithm --
 without any direction from its developers -- 
 has rederived one of the
 most fundamental aspects of physical oceanography.
 It has learned central features of the ocean
 from the patterns of \ssta\ alone.  
 In this regard, \ulmo\ may hold greater potential/promise
 to learn and derive other, not-yet-identified behaviors
 in the ocean.

\newpage
\subsection{Scrutinizing examples of the outliers}
\label{sec:scrutinize}

\begin{figure}[]
\centering
\includegraphics[width=15 cm]{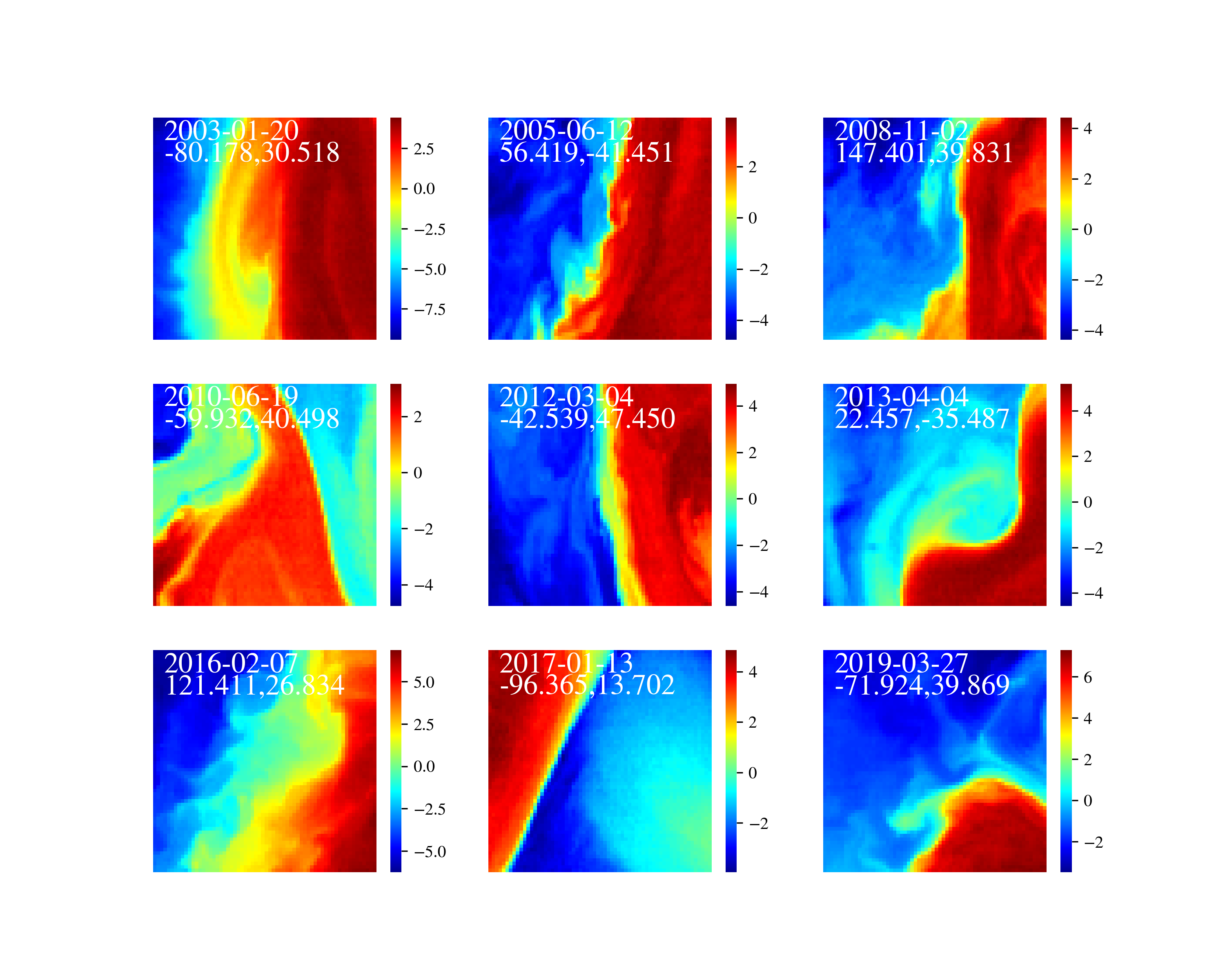}
\vskip -0.4in
\caption{Gallery of outliers drawn from the 
distribution of {\image}s with
the lowest 0.1\%\ of LL values.  
This representative sample was
further selected to uniformly sample the $\sim 20$~year interval of
the dataset and the spatial distribution
(Figure~\ref{fig:OOD_spatial}).
Note the large and complex temperature gradients within all {\image}s
showing $\mDT > 7$\,K.  The color bar refers to \ssta\ in units of Kelvin.
}
\label{fig:gallery_SSTa}
\end{figure}

Figure~\ref{fig:gallery_SSTa} shows a gallery 
of 9~outliers selected to uniformly
span time and location in the ocean.
These exhibit extreme {\ssta} variations and/or 
complexity and (presumably) mark significant
mesoscale activity.
A common characteristic of these {\image}s is
the presence of a strong and sharp 
gradient in {\ssta} which separates two regions exhibiting a 
large temperature difference.
Typically, such gradients are  associated with strong ocean currents, often at mid-latitudes on the western edge of ocean basins.
We define a simple statistic of the temperature distribution
$\mDT \equiv T_{90} - T_{10}$
where $T_X$ is the temperature at the Xth percentile of 
a given \image.
All of the outliers in Figure~\ref{fig:gallery_SSTa}
exhibit $\mDT > 7$K, a point we return to in the
following sub-section.


\begin{figure}[h]
\centering
\includegraphics[width=14cm]{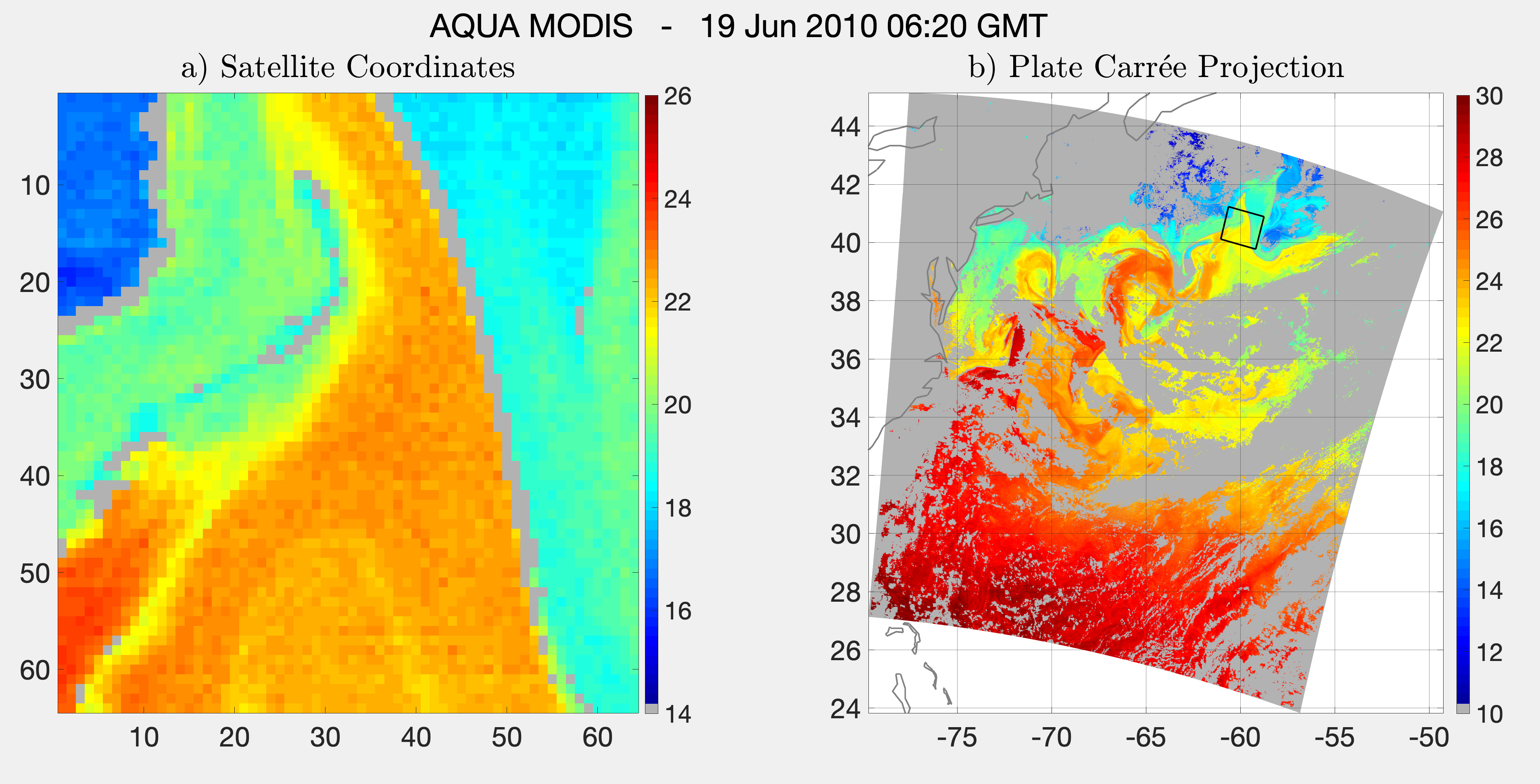}
\caption{The 19 June 2010 {\image} of Fig.~\ref{fig:gallery_SSTa} with \ac{LL} $=-1234.1$, in the lowest 0.1\% of the log-likelihood values for the dataset. 
a) The original {\image} in satellite-coordinates prior to preprocessing. b) The entire granule shown in a plate carr\'ee projection with the continental boundary (dark gray) and the {\image} (black rectangle). 
Note the change of palettes from a to b to accommodate the larger range of {\sst} in b. 
Light gray are masked pixels, either from clouds, land, or
clear pixels improperly flagged as cloud-contaminated 
($\S$~\ref{sec:data}). 
}
\label{fig:Gulf_Stream_Outlier}
\end{figure}

\begin{figure}[h]
\centering
\includegraphics[width=12 cm]{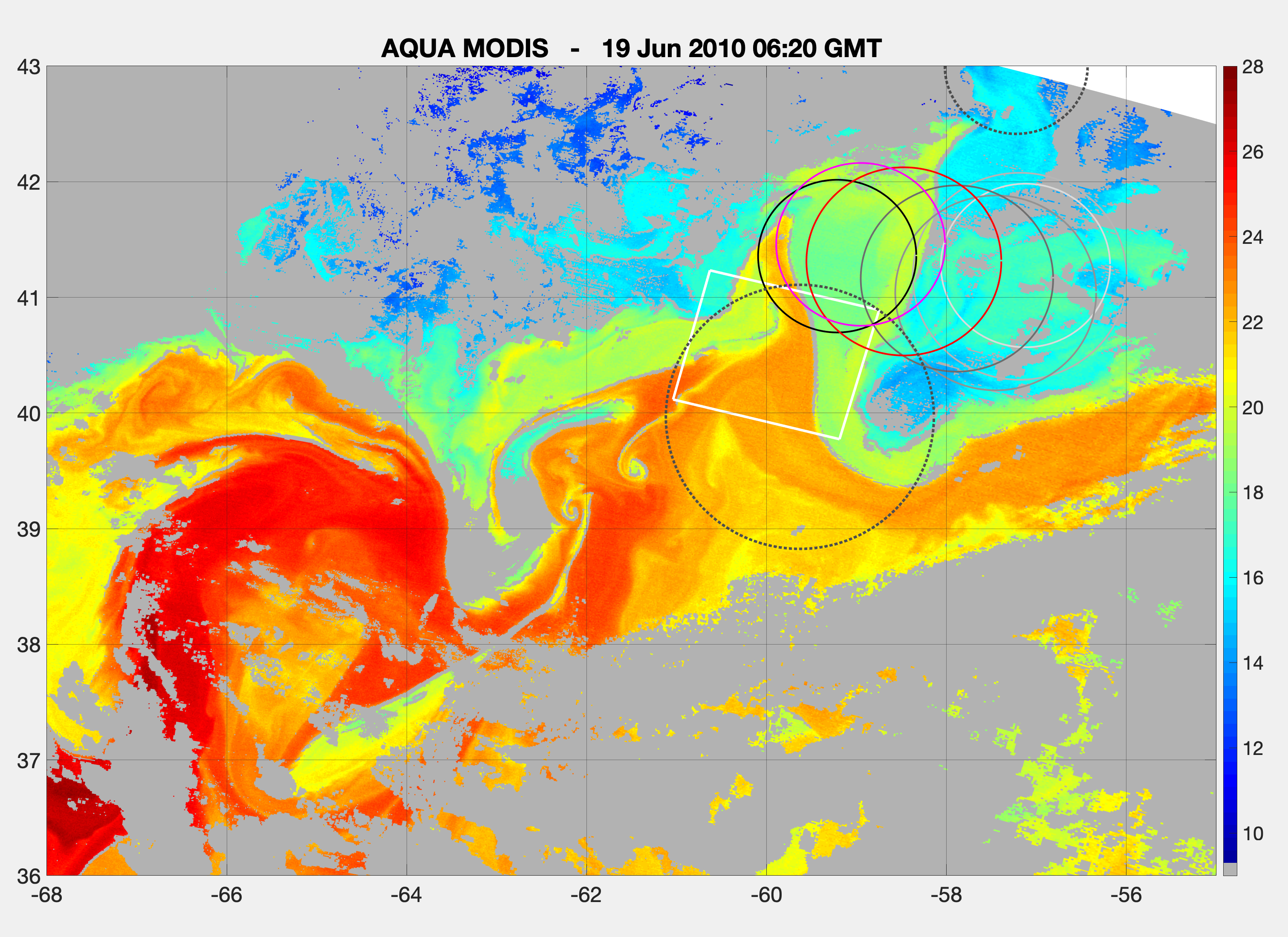}
\caption{The \ac{SST} field in the vicinity of the 19 June 2010 {\image} of Fig.~\ref{fig:gallery_SSTa}. The {\image} (white rectangle)
is centered on 60$^\circ$W, 41$^\circ$30"N. 
Solid circles outline all observations of the anticyclonic eddy identified by \citet{CHELTON2011167}, which passed through the {\image} on 19 June 2010. Circles are colored from light gray ($1^{st}$ detection of the eddy on 17 May) to black (last observation on 28 June) with the exception of its location shown in red on 14 June, immediately prior to the satellite overpass, and its location (magenta) one week later, 21 June, immediately following the overpass. 
Circles with dotted black outlines are anticyclonic eddies found in a broader search ($\pm3^\circ$ centered on 60$^\circ$W, 41$^\circ$30"N) on 5 July, one week after the last observation of the eddy of interest. No cyclonic eddies were found in a similar search. 
Clouds are light gray.
}
\label{fig:Gulf_Stream_Outlier_Details}
\end{figure}

As an example of the anomalous behaviour associated with \outliers, 
we examine the evolution of the {\sst} field in the vicinity of the 19 June 2010 {\image} (Fig.~\ref{fig:gallery_SSTa}) located in the Gulf Stream region; Fig.~\ref{fig:Gulf_Stream_Outlier}a shows the {\image} and (b) its location in the 5-minute granule. We selected this {\image} because it is in a region with which we have significant experience. Fig.~\ref{fig:Gulf_Stream_Outlier_Details} shows an expanded version of the {\sst} field in the vicinity of the {\image}. The main feature in Fig.~\ref{fig:Gulf_Stream_Outlier_Details} is the Gulf Stream, the bright red, fading to orange, band meandering from the bottom left hand corner of the image to the middle of the right hand side. A portion of the Gulf Stream loops through the lower half of the {\image} and a streamer extends to the north (Fig.~\ref{fig:Gulf_Stream_Outlier}) from the northernmost excursion of the stream. To aid in the interpretation of this {\image}, we make use of the mesoscale eddy dataset produced by \citet{CHELTON2011167}. It shows an eddy, most probably a \ac{WCR}, moving to the west at approximately 5\,km/day to the north of the stream from 17 May (very light gray circle) to 14 June (red circle) when it began to interact with the Gulf Stream, drawing warm Gulf Stream Water on its western side to the north and cold Slope Water on its eastern side to the south. 
The eddy disappears from the altimeter record two weeks later and is replaced by a very large anticyclone (the dotted black circle) to the west southwest of the eddy's last position. This is likely a detaching meander of the Gulf Stream resulting from the absorption of the eddy into an already chaotic configuration. 

Of particular interest is that the Gulf Stream appears to have lost its coherence between approximately 63$^\circ$ and 59$^\circ$W.  Specifically, note the very thin band of cooler water ($\sim 21^{\circ}$C) in the middle of the warm band ($\sim24^{\circ}$C) of, presumably, Gulf Stream Water between $63.5^{\circ}$ and $62^{\circ}$W and a second similar band (but moving in the opposite direction) between $61.5^{\circ}$ and $60.5^{\circ}$. The western cool band appears to separate one branch of Gulf Stream Water that has been advected from the southwestern edge of the large meander centered at $65.5^{\circ}$W, $38.5^{\circ}$N, and a second branch advected from its southeastern edge. 
These two branches may result from a general instability of the Gulf Stream associated with formation, or in this case the likely aborted formation, of a \ac{WCR}. In the normal formation process, the initial state is a large meander of the Gulf Stream and the final state is a relatively straight Gulf Stream with a \ac{WCR} to the north. In this case the process appears to have begun but inspection of subsequent images suggests that a ring was not formed; the meander reformed after initially beginning the detachment process. However, this is all quite speculative, the important point is that the stream appears to have lost its coherence immediately upstream of the cutout, which we believe to be a very unusual process.
Admittedly the {\image} only `sees' a very small portion of this but we have found the suggestion of convoluted dynamics in the immediate vicinity of a large fraction of other outliers
as well. Bottom line: Cornillon, who has been looking at {\sst} fields derived from satellite-borne sensors for over 40 years, found that more than one-in-ten of the anomalous fields discovered by ULMO suggested intriguing dynamics that he has not previously encountered; recall that this is one-in-ten of one-in-a-thousand (the definition of an outlier) or approximately one field in ten thousand.

\subsection{Digging Deeper}
\label{sec:digging}

It is evident from the preceding sub-sections 
(e.g., Figure~\ref{fig:gallery_SSTa}) that \ulmo\ has 
discovered a set of highly unusual and dynamic regions
of the ocean.  Scientifically, this is extremely useful 
-- irrespective of the underlying processes --
as it can launch future, deeper inquiry into the physical
processes generating such patterns.  On the other hand,
as scientists we are inherently driven to understand -- as
best as possible -- what/how/why \ulmo\ triggered upon.
We begin that process here and defer further exploration
to future work.

\begin{figure}[h]
\centering
\includegraphics[width=15 cm]{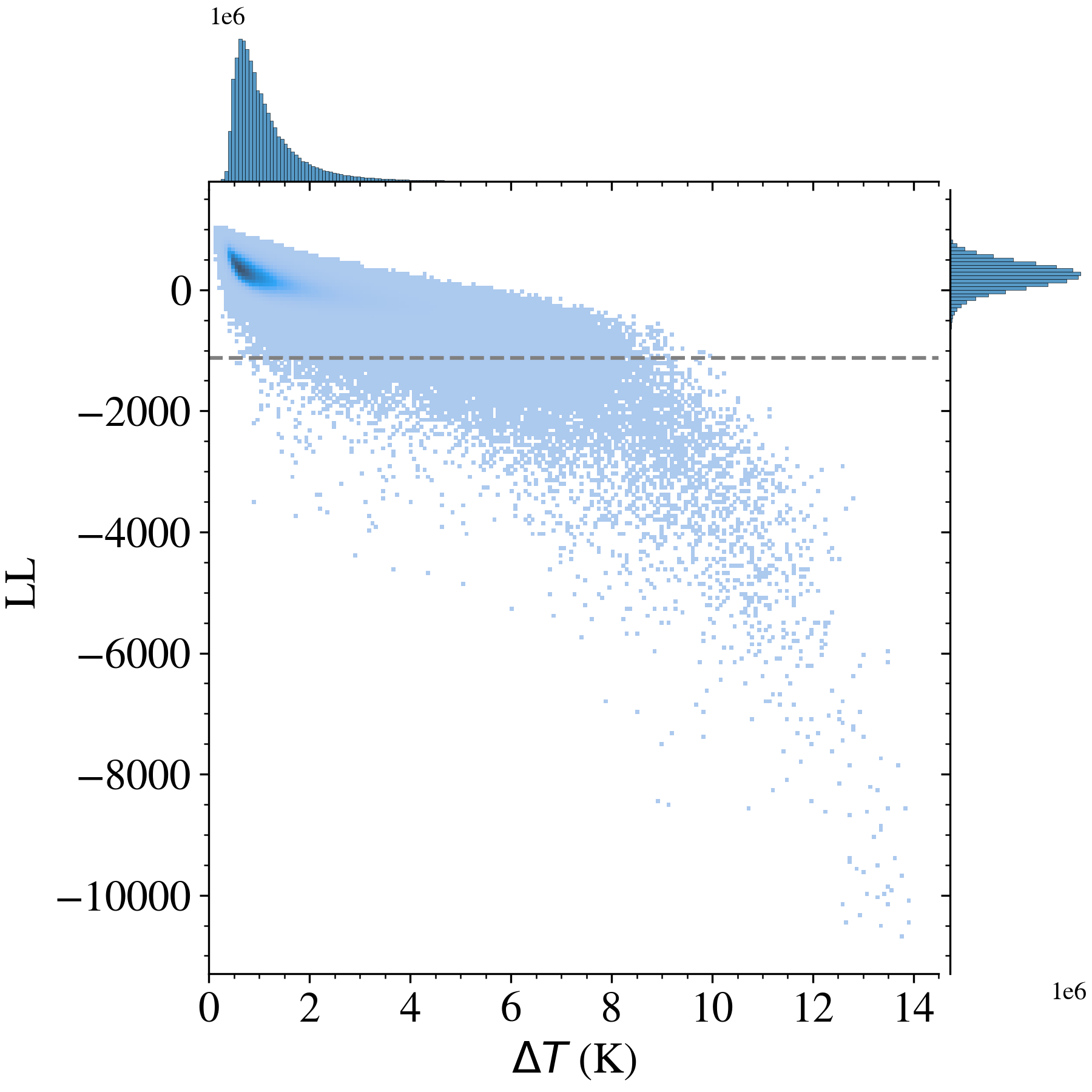}
\caption{Distribution of \ac{LL} values as a function of \DT.
While there is a strong anti-correlation apparent, the relationship
exhibits substantial scatter such that \DT\ is not a precise
predictor of \ac{LL} nor  
the underlying \ssta\ patterns characterized by \ulmo. 
The horizontal line at $\rm LL \approx -1123$ 
corresponds to the 0.1\% threshold; 
{\image}s with log-likelihood values beneath this line 
are considered to be outliers. 
}
\label{fig:LLvsDT}
\end{figure}

In Section~\ref{sec:anomalies}, we emphasized that the entire gallery
of outliers (Figure~\ref{fig:gallery_SSTa}) exhibits a large
temperature variation $\mDT > 7$K.
Exploring this further, Figure~\ref{fig:LLvsDT} plots 
LL vs.\ \DT\ for the full set of {\image}s analyzed.
Indeed, the two are anti-correlated with the lowest 
LL values corresponding to the largest \DT.
This suggests that a simple rules-based algorithm of selecting
all {\image}s with $\mDT > 10$\,K would select the most
extreme outliers discovered by \ulmo.  One may question,
therefore, whether a complex and hard-to-penetrate
AI model was even necessary to reproduce our results.

Further analysis suggests that there may be more to the distribution of \acp{LL}. Specifically, note that there is substantial scatter about the mean 
relation between \ac{LL} and \DT; for example, at LL~$= -1000$ one finds 
\DT\ values ranging from $1-10$\,K.
Similarly, 
any \image\ with $\mDT < 8$\,K includes a non-negligible
set of images with ${\rm LL} \gtrsim 0$.
Figure~\ref{fig:LLvsDT} indicates that the
patterns that \ulmo\ flags as outliers
are not solely determined by \DT.

\begin{figure}[h]
\centering
\includegraphics[width=16 cm]{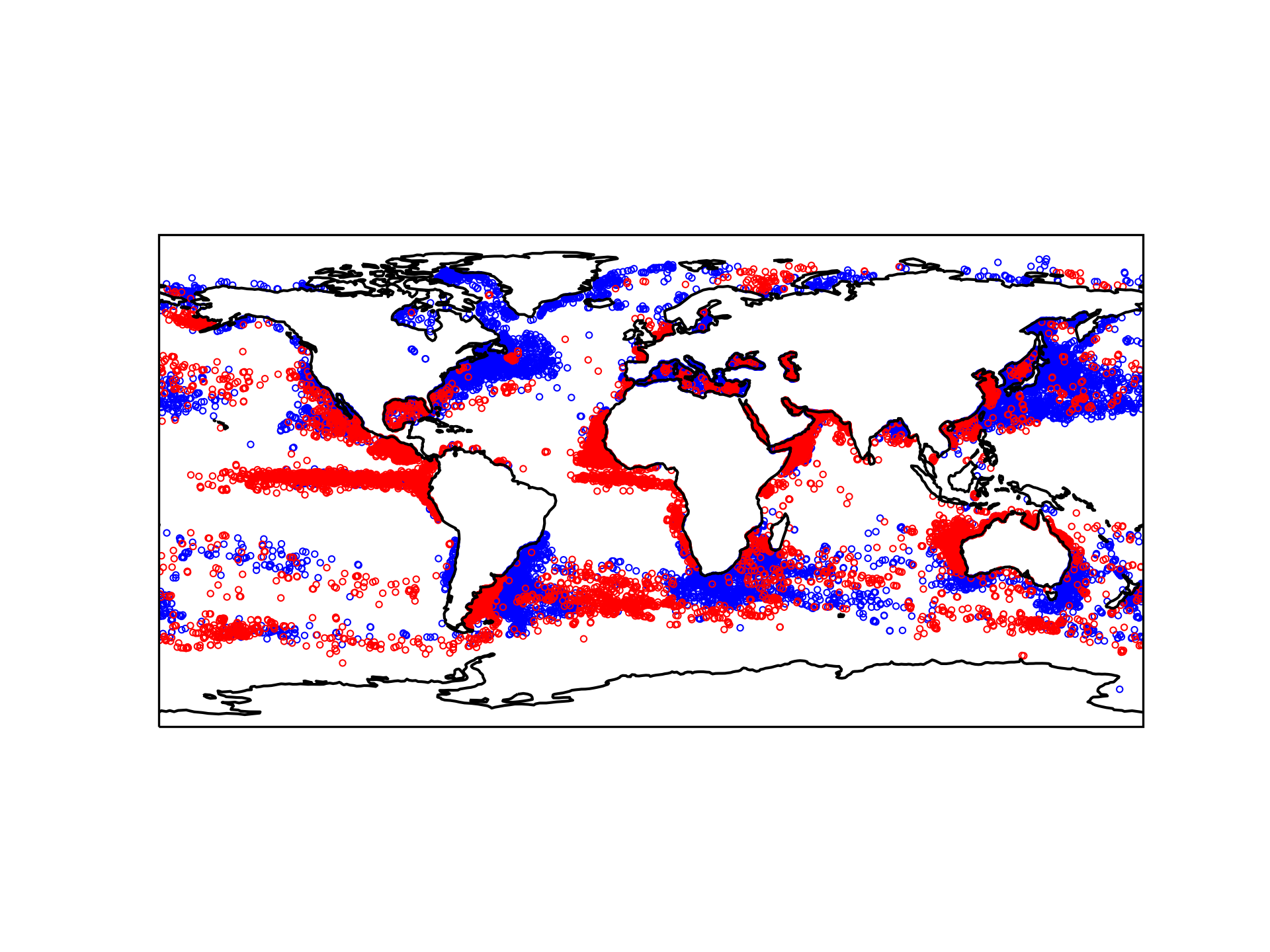}
\vskip -0.9in
\caption{
Spatial distribution of the {\LLten}/{\LLninetyCutouts}  
(red/blue)
defined as the upper/lower tenth percentile of the 
LL distribution for the set of {\image}s
with $\mDT = [2,2.1]$\,K.  It is evident that these {\image}s
occupy distinct regions of the ocean, 
i.e., the \ulmo\ algorithm
has identified patterns with significant spatial coherence.
More remarkable, note the several areas
(e.g.\ in the Brazil-Malvinas current) where one identifies
adjacent but separate patches of 
{\LLninety} and {\LLtenCutouts}.
}
\label{fig:T2_lowhiLL}
\end{figure}

This becomes especially clear in the following exercise.
Consider the full set of {\image}s within the small range 
$\mDT = [2-2.1]$\,K. 
From Figure~\ref{fig:LLvsDT}, we see
these exhibit ${\rm LL} \approx [-2600,590]$
and find that the LL distribution 
is well described by a Gaussian (not shown) with 
$<{\rm LL}> \approx 10$ and $\sigma({\rm LL}) \approx 150$.
Now consider the {\image}s with the lowest/highest 10/90\%\ 
of the distribution, i.e., the 'outlier'/'inlier'
sub-samples within this small range of \DT. 
We refer to these as $LL_{10}$ and $LL_{90}$ cutouts, respectively. 
Figure~\ref{fig:T2_lowhiLL} shows the spatial distribution
of these {\image}s. 
Remarkably, there are multiple areas dominated by only one of the
sub-samples (e.g., {\LLninetyCutouts} along the Pacific equator).
It is evident that \ulmo\ finds large spatial structures in the log-likelihood distribution of cutouts that are {\it independent} of $\Delta T$.

\begin{figure}[h]
\centering
\includegraphics[width=15 cm]{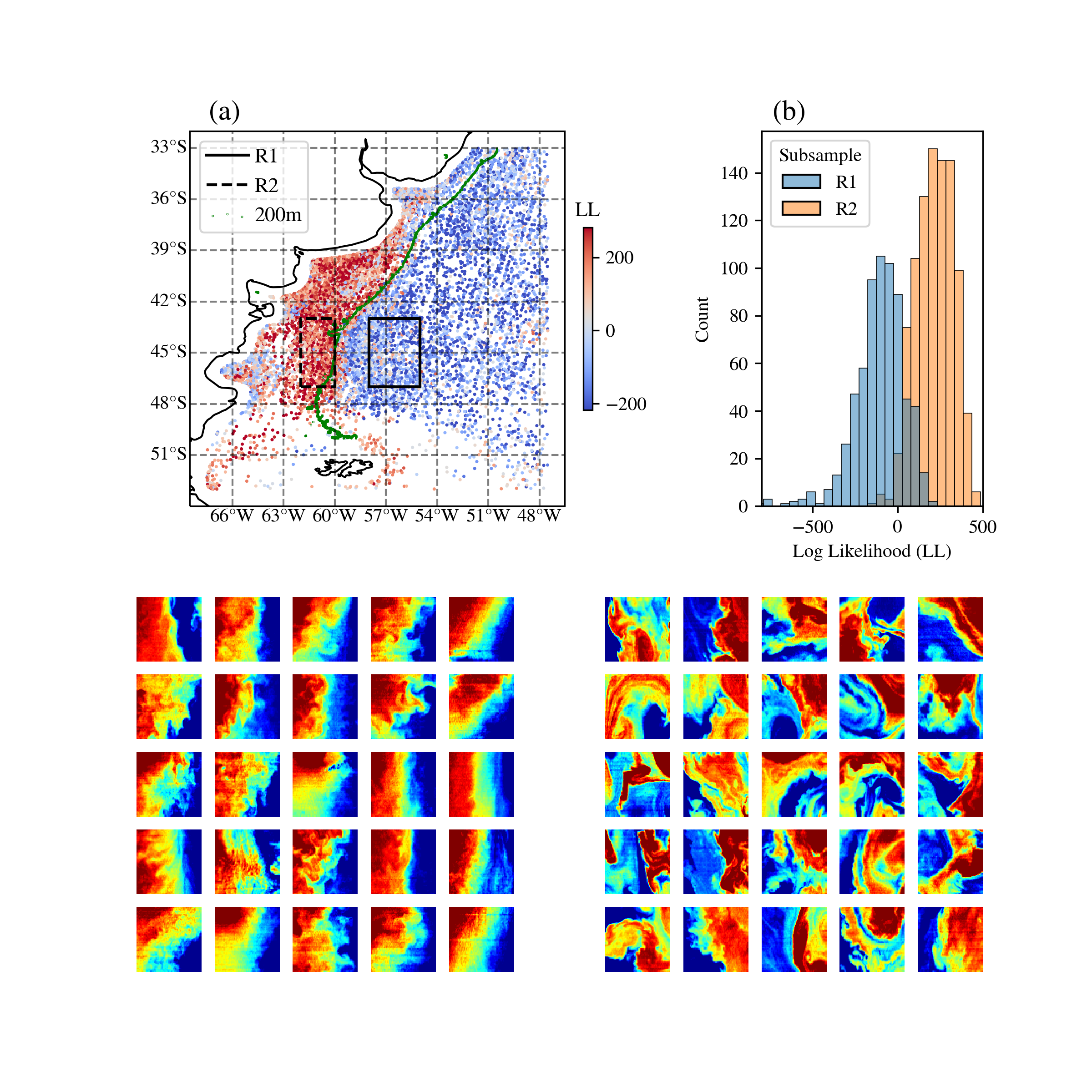}
\vskip -0.6in
\caption{
(a) Distribution of the \ac{LL} values
for cutouts near the  Brazil-Malvinas Confluence, 
restricted to those with temperature
difference $\mDT \approx 2$\,K.
One identifies
a clear separation, where the lower \ac{LL} values
lie within the current and the higher values
lie close to the Argentinian coast.
Marked are two rectangles (R1, R2), one in
each region, referred to in the other panels.
Also marked is the 200m bathymetry described
in the text.
(b)
Histograms of the \ac{LL} values for the 
{\image}s from two regions (R1/R2) chosen to show
lower/higher \ac{LL} values near the confluence.
(lower panels) Representative cutouts from each subset --
the left set are drawn from the R2 rectangle and therefore 
exhibit higher \ac{LL} values.  The right set are from R1.
These galleries reveal qualitative differences in 
the SST patterns, i.e., unique ocean dynamics. 
The palette for the lower panels ranges linearly from $-1$ to 1\,K.
}
\label{fig:Argentina}
\end{figure}

Furthermore,  there are 
several locations in the ocean where
{\LLten} and {\LLninetyCutouts} are adjacent to one another but still separate.
One clear example is within the Brazil-Malvinas Confluence, 
off the coast of Argentina.
Figure~\ref{fig:Argentina}a  shows a zoom-in of that region with the colors corresponding to the \ac{LL} values 
(not strictly the {\LLten} or {\LLninety} distributions shown in Fig.~\ref{fig:T2_lowhiLL}). 
Figure~\ref{fig:Argentina}a highlights the clear and striking separation of the \ac{LL} values in this region as do the histograms (Figure~\ref{fig:Argentina}b) for the \ac{LL} values of cutouts in the two rectangles shown in panel a.
The dynamics of the ocean in this region is well-studied \cite{Piola:2018hd}.
Higher \ac{LL} regions tend to be 
found on the 
Patagonian Shelf where the dynamics are 
dominated by 
tides, buoyancy and wind--forcing the circulation at the local level--and off-shore currents--forcing the circulation remotely.
In contrast, the lower \ac{LL} regions track more dynamic, current-driven
motions of the main Brazil-Malvinas Confluence. 
Of particular interest is the rather abrupt switch at $\sim40^{\circ}$S from higher \ac{LL} values to the south to lower values 
to the north. 
This is consistent with the observation of \citet{Combes:2018dg}, based on numerical simulations, that ``[t]here is an abrupt change of the dynamical characteristics of the shelf circulation at 40$^{\circ}$S''. They attribute this change in dynamics to this region being a sink for Patagonian Shelf waters, which are being advected offshore by the confluence of the Brazil and Malvinas Currents. Again, {\ulmo} has captured striking detail in regional dynamics with no directed input. Further analysis of the region (not shown) suggests that {\ulmo} has also captured seasonal differences in the dynamics, with a region of lower \ac{LL} cutouts in waters approximately 100\,m deep between $38^{\circ}$ and $45^{\circ}$S in austral winter but not austral summer.

Intrigued by \ulmo's ability to spatially separate these
regions based on \ssta\ patterns alone, we 
inspected a set of 25 randomly selected samples from R1, the eastern rectangle in Figure~\ref{fig:Argentina}a, and 25 randomly selected samples from R2 
to further explore its inner-workings
(see lower panels of Figure~\ref{fig:Argentina}).
The comparison is striking and we easily identify
qualitative differences in the observed patterns despite
their nearly identical \DT\ values.
The higher \ac{LL} cutouts show large-scale gradients and features with significant coherence
whereas the lower \ac{LL} cutouts exhibit gradients and features with
a broader range of scales and a suggested richer distribution of relative
vorticity. 

Another area, which stands out in Figure~\ref{fig:T2_lowhiLL}, is that in the Northwest Atlantic where a region of {\LLninetyCutouts} (red) are surrounded by {\LLtenCutouts} (blue). The structure (not shown) of the {\LLninetyCutouts} in this region, which are on the Grand Banks of Newfoundland, resemble the structure of the {\LLninetyCutouts} shown in Figure~\ref{fig:Argentina} and the structure of the {\LLtenCutouts} in this region is much closer to that of the {\LLtenCutouts} shown in Figure~\ref{fig:Argentina} than to the {\LLninetyCutouts} in either region. In fact, a gallery of randomly selected {\LLninetyCutouts} from the world ocean are similar to those off of Argentina and Newfoundland and a gallery of randomly selected {\LLtenCutouts} from the world ocean are more similar to the {\LLtenCutouts} off of Argentina and Newfoundland than to the {\LLninetyCutouts}. Simply put, the structure of the {\sst} cutouts shown in blue in Figure~\ref{fig:T2_lowhiLL} tend to be similar to one another and quite different from those shown in red although the cutouts in both cases have virtually the same dynamic ranges in {\sst}. This observation raises intriguing questions about the similarities and the differences in upper ocean processes in these regions -- questions to be addressed in further analyses of the fields.

We can capture some of the differences between the higher/lower \ac{LL} sub-samples 
of Figure~\ref{fig:Argentina} with another simple
statistic -- the RMS in SSTa, \sT.  On average, the lower \ac{LL} cutouts
exhibit 
$\approx 9\%$ higher \sT\ than 
those with a higher \ac{LL}.
Furthermore, we find LL correlates with \sT\ in a fashion similar to \DT. 
On the other hand, it is evident from Figure~\ref{fig:Argentina}
that there is significant structure apparent in the {\image}s that
is not described solely by \sT. 
The correlations of \ac{LL} with \DT\ and \sT\ manifest from the 
underpinnings of \ulmo: the distribution of ocean \ssta\ patterns reflect
the distribution of
simple statistics like \DT\ or \sT, which exhibit large and
non-uniform variations across the ocean.
The complexity of these patterns, however, belie the information
provided by simple statistics alone.

\section{Conclusions and Future Work}
\label{sec:conclusions}

With the design and application of a machine learning 
algorithm \ulmo, we
set out to identify the rarest sea surface temperature patterns
in the ocean 
through an out-of-distribution analysis yielding a
unique log-likelihood (\ac{LL}) value for every \image.  
On this goal, we believe we were successful
(e.g., Figures~\ref{fig:OOD_spatial},\ref{fig:gallery_SSTa}).
In examining the nature of the outliers 
we found that these exhibited extrema of two
simple metrics: the temperature difference \DT\ and 
standard deviation \sT.
With the full privilege of hindsight, we
expect that any metric introduced to describe
the cutouts which exhibits a broad and non-uniform
distribution would correlate with \ac{LL}.
However, no single metric can capture the inherent pattern 
complexity and therefore none correlates tightly
with \ac{LL} (Figure~\ref{fig:LLvsDT}).

Looking to the future, the greatest potential
of algorithms like \ulmo\ may be that the patterns it learns
are more fundamental than 
measures traditionally implemented in the
scientific community (e.g., \ac{FFT}, Empirical Orthogonal Function \ac{EOF}).  
We hypothesize that the mathematical nature of 
convolutional neural networks \ac{CNN} 
-- convolutional features and max-pooling, which
synthesizes data across the scene while remaining
invariant to translation --
captures aspects of the data
that EOF analysis could not (nor any other simple
linear approach).  
Indeed, referring back
to Figure~\ref{fig:Argentina}, while as humans
we trivially distinguish between the two sets
of cutouts marking the ocean dynamics in the 
Brazil-Malvinas Confluence and can identify 
metrics on which they differ, these
metrics offer incomplete descriptions.  Going
forward, we will determine the extent 
(e.g., via analysis of ocean model outputs)
to which the patterns mark fundamental, dynamical
processes within the ocean. 
Potentially, the patterns learned by \ulmo\ 
(or its successors) hold the optimal description
of any such phenomena.

As emphasized at the onset, this manuscript offers
only a first glimpse at the potential for applying
advanced artificial intelligence techniques to the tremendous ocean
datasets obtained from satellite-borne sensors.
The techniques introduced
here will translate seamlessly to sea surface height
or ocean color imaging to
identify extrema/complexity in geostrophic currents and 
biogeochemical processes.  These too will
be the focus of future works.

\bigskip

\centerline{ACRONYMS}
\begin{doublespace}
\begin{acronym}[12345678901234]
\acro{(A)ATSR}{one or all of \textsmaller{ATSR}, \textsmaller{ATSR-2} and \textsmaller{AATSR}}
\acro{AATSR}{Advanced Along Track Scanning Radiometer}
\acro{ACC}{Antarctic Circumpolar Current}
\acro{ACCESS}{Advancing Collaborative Connections for Earth System Science}
\acro{ACL}{Access Control List}
\acro{ACSPO}{Advanced Clear Sky Processor for Oceans}
\acro{ADA}{Automatic Detection Algorithm}
\acro{ADCP}{Acoustic Doppler Current Profiler}
\acro{ADT}{absolute dynamic topography}
\acro{AESOP}{Assessing the Effects of Submesoscale Ocean Parameterizations}
\acro{AGU}{American Geophysical Union}
\acro{AI}{Artificial Intelligence}
\acro{AIRS}{Atmospheric Infrared Sounder}
\acro{AIS}{Ancillary Information Service}
\acro{AIST}{Advanced Information Systems Techonology}
\acro{AISR}{Applied Information Systems Research}
\acro{ADL}{Alexandria Digital Library}
\acro{API}{Application Program Interface}
\acro{APL}{Applied Physics Laboratory}
\acro{API}{Application Program Interface}
\acro{AMSR}{Advanced Microwave Scanning Radiometer}
\acro{AMSR2}{Advanced Microwave Scanning Radiometer 2}
\acro{AMSR-E}{Advanced Microwave Scanning Radiometer - \textsmaller{EOS}}
\acro{ANN}{ Artificial Neural Network}
\acro{AOOS}{Alaska Ocean Observing System}
\acro{APAC}{Australian Partnership for Advanced Computing}
\acro{APDRC}{Asia-Pacific Data-Research Center}
\acro{ARC}{{\smaller ATSR} Reprocessing for Climate}
\acro{ASCII}{American Standard Code for Information Interchange}
\acro{AS}{Aggregation Server}
\acro{ASFA}{Aquatic Sciences and Fisheries Abstracts}
\acro{ASTER}{Advanced Spaceborne Thermal Emission and Reflection Radiometer}
\acro{ATBD}{Algorithm Theoretical Basis Document}
\acro{ATSR}{Along Track Scanning Radiometer}
\acro{ATSR-2}{Second \textsmaller{ATSR}}
\acro{AVISO}{Archiving, Validation and Interpretation of Satellite Oceanographic Data}
\acro{ANU}{Australian National University}
\acro{AVHRR}{Advanced Very High Resolution Radiometer}
\acro{AzC}{Azores Current}

\acro{BAA}{Broad Agency Announcement}
\acro{BAO}{bi-annual oscillation}
\acro{BES}{Back-End Server}
\acro{BMRC}{Bureau of Meteorology Research Centre}
\acro{BOM}{Bureau of Meteorology}
\acro{BT}{brightness temperature}
\acro{BUFR}{Binary Universal Format Representation}

\acro{CAN}{Cooperative Agreement Notice}
\acro{CAS}{Community Authorization Service}
\acro{CC}{cloud cover}
\acro{CCA}{Cayula-Cornillon Algoritm}
\acro{CCI}{Climate Change Initiative}
\acro{CCLRC}{Council for the Central Laboratory of the Research Councils}
\acro{CCMA}{Center for Coastal Monitoring and Assessment}
\acro{CCR}{cold core ring}
\acro{CCS}{California Current System }
\acro{CCSM}{Community Climate System Model}
\acro{CCSR}{Center for Climate System Research}
\acro{CCV}{Center for Computation and Visualization}
\acro{CDAT}{Climate Data Analysis Tools}
\acro{CDC}{Climate Diagnostics Center}
\acro{CDF}{Common Data Format}
\acro{CDR}{Common Data Representation}
\acro{CEDAR}{Coupled Energetic and Dynamics and Atmospheric Regions}
\acro{CEOS}{Committee on Earth Observation Satellites}
\acro{CERT}{Computer Emergency Response Team}
\acro{CenCOOS}{Central \& Northern California Ocean Observing System}
\acro{CF}{clear fraction}
\acro{CGI}{Common Gateway Interface}
\acro{CHAP}{\textsmaller{CISL} High Performance Computing Advisory Panel}
\acro{CIFS}{Common Internet File System}
\acro{CIMSS}{Cooperative Institute for Meteorological Satellite Studies}
\acro{CIRES}{Cooperative Institute for Research (in) Environmental Sciences}
\acro{CISL}{Computational \& Information Systems Laboratory}
\acro{CLASS}{Comprehensive Large Array-data Stewardship System}
\acro{CLIVAR}{Climate Variability and Predictability}
\acro{CLS}{Collecte Localisation Satellites}
\acro{CME}{Community Modeling Effort}
\acro{CMS}{Centre de M\'et\'eorologie Spatiale}
\acro{CNN}{Convolutional Neural Network}
\acro{COA}{Climate Observations and Analysis}
\acro{COARDS}{Cooperative Ocean-Atmosphere Research Data Standard}
\acro{COAPS}{Center for Ocean-Atmospheric Prediction Studies}
\acro{COBIT}{Control Objectives for Information and related Technology}
\acro{COCO}{{\smaller CCSR} Ocean Component model}
\acro{CODAR}{Coastal Ocean Dynamics Applications Radar}
\acro{CODMAC}{Committee on Data Management, Archiving, and Computing}
\acro{Co-I}{Co-Investigator}
\acro{CORBA}{Common Object Request Broker Architecture}
\acro{COLA}{Center for Ocean-Land-Atmosphere Studies}
\acro{CPU}{Central Processor Unit}
\acro{CRS}{Coordinate Reference System}
\acro{CSA}{Cambridge Scientific Abstracts}
\acro{CSC}{Coastal Services Center}
\acro{CSIS}{Center for Strategic and International Studies}
\acro{CSL}{Constraint Specification Language}
\acro{CSP}{Chermayeff, Sollogub and Poole, Inc.}
\acro{CSDGM}{Content Standard for Digital Geospatial Metadata}
\acro{CSV}{Comma Separated Values}
\acro{CTD}{Conductivity, Temperature and Salinity}
\acro{CVSS}{Common Vulnerability Scoring System}
\acro{CZCS}{Coastal Zone Color Scanner}

\acro{DAAC}{Distribute Active Archive Center}
\acro{DAARWG}{Data Archiving and Access Requirements Working Group}
\acro{DAP}{Data Access Protocol}
\acro{DAS}{Data set Attribute Structure}
\acro{DBMS}{Data Base Management System}
\acro{DBDB2}{Digital Bathymetric Data Base} 
\acro{DChart}{Dapper Data Viewer}
\acro{DDS}{Data Descriptor Structure}
\acro{DDX}{\textsmaller{XML} version of the combined \textsmaller{DAS} and \textsmaller{DDS}}
\acro{DFT}{Discrete Fourier Transform}
\acro{DIF}{Directory Interchange Format}
\acro{DISC}{Data and Information Services Center}
\acro{DIMES}{Diapycnal and Isopycnal Mixing Experiment:  Southern Ocean}
\acro{DMAC}{Data Management and Communications committee}
\acro{DMR}{Department of Marine Resources}
\acro{DMSP}{Defense Meteorological Satellite Program}
\acro{DoD}{Department of Defense}
\acro{DODS}{Distributed Oceanographic Data System}
\acro{DOE}{Department of Energy}
\acro{DSP}{U. Miami satellite data processing software}
\acro{DSS}{direct statistical simulation}

\acro{EASy}{Environmental Analysis System}
\acro{ECCO}{Estimating the Circulation and Climate of the Ocean}
\acro{ECCO2}{{\smaller  Estimating the Circulation and Climate of the Ocean} Phase II}
\acro{ECS}{\textsmaller{EOSDIS} Core System}
\acro{ECHO}{Earth Observing System Clearinghouse}
\acro{ECMWF}{European Centre for Medium-range Weather Forecasting}
\acro{ECV}{Essential Climate Variable}
\acro{EDC}{Environmental Data Connector}
\acro{EDJ}{Equatorial Deep Jet}
\acro{EDFT}{Extended Discrete Fourier Transform}
\acro{EDMI}{Earth Data Multi-media Instrument}
\acro{EEJ}{Extra-Equatorial Jet}
\acro{EIC}{Equatorial Intermediate Current}
\acro{EICS}{Equatorial Intermediate Current System}
\acro{EJ}{Equatorial Jets}
\acro{EKE}{eddy kinetic energy}
\acro{EMD}{Empirical Mode Decomposition}
\acro{EOF}{Empirical Orthogonal Function}
\acro{EOS}{Earth Observing System}
\acro{EOSDIS}{Earth Observing System Data Information System}
\acro{EPA}{Environmental Protection Agency}
\acro{EPSCoR}{Experimental Program to Stimulate Competitive Research}
\acro{EPR}{East Pacific Rise}
\acro{ERD}{Environmental Research Division}
\acro{ERS}{European Remote-sensing Satellite}
\acro{ESA}{European Space Agency}
\acro{ESDS}{Earth Science Data Systems}
\acro{ESDSWG}{Earth Science Data Systems Workign Group}
\acro{ESE}{Earth Science Enterprise}
\acro{ESG}{Earth System Grid}
\acro{ESG II}{Earth System Grid -- II}
\acro{ESIP}{Earth Science Information Partner}
\acro{ESMF}{Earth System Modeling Framework}
\acro{ESML}{Earth System Markup Language}
\acro{ESP}{eastern South Pacfic}
\acro{ESRI}{Environmental Systems Research Institute}
\acro{ESR}{Earth and Space Research}
\acro{ETOPO}{Earth Topography}
\acro{EUC}{Equatorial Undercurrent}
\acro{EUMETSAT}{European Organisation for the Exploitation of Meteorological Satellites}
\acro{Ferret}{}

\acro{FASINEX}{Frontal Air-Sea Interaction Experiment}
\acro{FDS}{Ferret Data Server}
\acro{FFT}{Fast Fourier Transform}
\acro{FGDC}{Federal Geographic Data Committee}
\acro{FITS}{Flexible Image (or Interchange) Transport System}
\acro{FLOPS}{FLoating point Operations Per Second} 
\acro{FRTG}{Flow Rate Task Group}
\acro{FreeForm}{}
\acro{FNMOC}{Fleet Numerical Meteorology and Oceanography Center}
\acro{FSU}{Florida State University}
\acro{FTE}{Full Time Equivalent}
\acro{ftp}[\normalsize  ftp]{File Transport Protocol}
\acro{FTP}[\normalsize  ftp]{File Transport Protocol}

\acro{GAC}{Global Area Coverage}
\acro{GAN}{Generative Adversarial Network}
\acro{GB}{GigaByte - $10^{9}$ bytes}
\acro{GCMD}{Global Change Master Directory}
\acro{GCM}{general circulation model}
\acro{GCOM-W1}{Global Change Observing Mission - Water}
\acro{GCOS}{Global Climate Observing System}
\acro{GDAC}{Global Data Assembly Center}
\acro{GDS}{\textsmaller{GrADS} Data Server}
\acro{GDS2}{GHRSST Data Processing Specification v2.0}
\acro{GEBCO}{General Bathymetric Charts of the Oceans}
\acro{GeoTIFF}{Georeferenced Tag Image File Format}
\acro{GEO-IDE}{Global Earth Observation Integrated Data Environment}
\acro{GES DIS}{Goddard Earth Sciences Data and Information Services Center}
\acro{GEMPACK}{General Equilibrium Modelling PACKage}
\acro{GEOSS}{Global Earth Observing System of Systems}
\acro{GFDL}{Geophysical Fluid Dynamics Laboratory}
\acro{GFD}{Geophysical Fluid Dynamics}
\acro{GHRSST}{Group for High Resolution Sea Surface Temperature}
\acro{GHRSST-PP}{\textsmaller{GODAE} High Resolution Sea Surface Temperature Pilot Project}
\acro{GINI}{\textsmaller{GOES} Ingest and \textsmaller{NOAA/PORT} Interface}
\acro{GIS}{Geographic Information Systems} 
\acro{Globus}{} 
\acro{GMAO}{Global Modeling and Assimilation Office}
\acro{GML}{Geography Markup Language}
\acro{GMT}{Generic Mapping Tool}
\acro{GODAE}{Global Ocean Data Assimilation Experiment} 
\acro{GOES}{Geostationary Operational Environmental Satellites} 
\acro{GOFS}{Global Ocean Forecasting System}
\acro{GoMOOS}{Gulf of Maine Ocean Observing System}
\acro{GOOS}{Global Ocean Observing System}
\acro{GOSUD}{Global Ocean Surface Underway Data}
\acro{GPFS}{ General Parallel File System}
\acro{GPU}{Graphics Processing Unit}
\acro{GRACE}{Gravity Recovery and Climate Experiment} 
\acro{GRIB}{GRid In Binary} 
\acro{GrADS}{Grid Analysis and Display System} 
\acro{GridFTP}{\textsmaller{FTP} with GRID enhancements}
\acro{GRIB}{GRid in Binary} 
\acro{GPS}{Global Positioning System}
\acro{GSFC}{Goddard Space Flight Center} 
\acro{GSI}{Grid Security Infrastructure}
\acro{GSO}{Graduate School of Oceanography}
\acro{GTSPP}{Global Temperature and Salinity Profile Program}
\acro{GUI}{Graphical User Interface}
\acro{GS}{Gulf Stream}

\acro{HAO}{High Altitude Observatory} 
\acro{HLCC}{Hawaiian Lee Countercurrent}
\acro{HCMM}{Heat Capacity Mapping Mission}
\acro{HDF}{Hierarchical Data Format}
\acro{HDF-EOS}{Hierarchical Data Format - \textsmaller{EOS}} 
\acro{HEC}{High-End Computing}
\acro{HF}{High Frequency}
\acro{HGE}{High Gradient Event}
\acro{HPC}{High Performance Computing}
\acro{HPCMP}{High Performance Computing Modernization Program}
\acro{HPSS}{High Performance Storage System} 
\acro{HR DDS}{High Resolution Diagnostic Data Set}
\acro{HRPT}{High Resolution Picture Transmission}
\acro{HTML}{Hyper Text Markup Language}
\acro{html}{Hyper Text Markup Language}
\acro{http}{the hypertext transport protocol}
\acro{HTTP}{Hyper Text Transfer Protocol}
\acro{HTTPS}{Secure Hyper Text Transfer Protocol} 
\acro{HYCOM}{HYbrid Coordinate Ocean Model} 

\acro{I-band}{imagery resolution band}
\acro{IDD}{Internet Data Distribution}
\acro{IB}{Image Band}
\acro{IBL}{internal boundary layer}
\acro{IBM}{Internation Business Machines}
\acro{ICCs}{Intermediate Countercurrents}
\acro{IDE}{Integrated Development Environment}
\acro{IDL}{Interactive Display Language}
\acro{IDLastro}{\textsmaller{IDL} Astronomy User's Library}
\acro{IDV}{Integrated Data Viewer}
\acro{IEA}{Integrated Ecosystem Assessment}
\acro{IEEE}{Institute (of) Electrical (and) Electronic Engineers}
\acro{IETF}{Internet Engineering Task Force}
\acro{IFREMER}{Institut Fran\c{c}ais de Recherche pour l'Exploitation de la MER}
\acro{IMAPRE}{El Instituto del Mar del Per\'u}
\acro{IMF}{Intrinsic Mode Function}
\acro{IOOS}{Integrated Ocean Observing System}
\acro{ISAR}{Infrared Sea surface temperature Autonomous Radiometer}
\acro{ISO}{International Organization for Standardization}
\acro{ISSTST}{Interim Sea Surface Temperature Science Team}
\acro{IT}{Information Technology}
\acro{ITCZ}{Intertropical Convergence Zone} 
\acro{IP}{Internet Provider}
\acro{IPCC}{Intergovernmental Panel on Climate Change}
\acro{IPRC}{International Pacific Research Center}
\acro{IR}{Infrared}
\acro{IRI}{International Research Institute for Climate and Society}
\acro{ISO}{International Standards Organization}

\acro{JASON}{JASON Foundation for Education}
\acro{JDBC}{Java Database Connectivity}
\acro{JFR}{Juan Fern\'andez Ridge}
\acro{JGOFS}{Joint Global Ocean Flux Experiment}
\acro{JHU}{Johns Hopkins University}
\acro{JPL}{Jet Propulsion Laboratory}
\acro{JPSS}{Joint Polar Satellite System}

\acro{KDE}{Kernel Density Estimation}
\acro{KVL}{Keyword-Value List}
\acro{KML}{Keyhole Markup Language}
\acro{KPP}{K-Profile Parameterization}

\acro{LAC}{Local Area Coverage}
\acro{LAN}{Local Area Network}
\acro{LAS}{Live Access Server}
\acro{LASCO}{Large Angle and Spectrometric Coronagraph Experiment}
\acro{LatMIX}{Scalable Lateral Mixing and Coherent Turbulence}
\acro{LDAP}{Lightweight Directory Access Protocol}
\acro{LDEO}{Lamont Doherty Earth Observatory}
\acro{LEAD}{Linked Environments for Atmospheric Discovery}
\acro{LEIC}{Lower Equatorial Intermediate Current}
\acro{LES}{Large Eddy Simulation}
\acro{L1}{Level-1}
\acro{L2}{Level-2}
\acro{L3}{Level-3}
\acro{L4}{Level-4}
\acro{LL}{log-likelihood}
\acro{LLC}{Latitude/Longitude/polar-Cap}
\acro{LLC4320}[LLC-4320]{\ac{LLC}-4320}
\acro{LLC2160}[LLC-2160]{\ac{LLC}-2160}
\acro{LLC1080}[LLC-1080]{\ac{LLC}-1080}
\acro{LHF}{Latent Heat Flux}
\acro{LST}{local sun time}
\acro{LTER}{Long Term Ecological Research Network}
\acro{LTSRF}{Long Term Stewardship and Reanalysis Facility}
\acro{LUT}{Look Up Table}

\acro{M-band}{moderate resolution band}
\acro{MABL}{marine atmospheric boundary layer}
\acro{MADT}{Maps of Absolute Dynamic Topography}
\acro{MapServer}{MapServer}
\acro{MAT}{Metadata Acquisition Toolkit}
\acro{MATLAB}{}
\acro{MARCOOS}{Mid-Atlantic Coastal Ocean Observing System}
\acro{MARCOORA}{Mid-Atlantic Coastal Ocean Observing Regional Association}
\acro{MB}{MegaByte - $10^{6}$ bytes}
\acro{MCC}{Maximum Cross-Correlation}
\acro{MCR}{\textsmaller{MATLAB} Component Runtime}
\acro{MCSST}{Multi-Channel Sea Surface Temperature}
\acro{MDT}{mean dynamic topography}
\acro{MDB}{Match-up Data Base}
\acro{MDOT}{mean dynamic ocean topography}
\acro{MEaSUREs}{Making Earth System data records for Use in Research Environments}
\acro{MERRA}{Modern Era Retrospective-Analysis for Research and Applications}
\acro{MERSEA}{Marine Environment and Security for the European Area}
\acro{MTF}{Modulation Transfer Function}
\acro{MICOM}{Miami Isopycnal Coordinate Ocean Model}
\acro{MIRAS}{Microwave Imaging Radiometer with Aperture Synthesis}
\acro{MITgcm}{{\smaller MIT} General Circulation Model}
\acro{MIT}{Massachusetts Institute of Technology}
\acro{mks}{meters, kilograms, seconds}
\acro{MLP}{Multilayer Perceptron}
\acro{MLSO}{Mauna Loa Solar Observatory}
\acro{MM5}{Mesoscale Model}
\acro{MMI}{Marine Metadata Initiative}
\acro{MMS}{Minerals Management Service}
\acro{MODAS}{Modular Ocean Data Assimilation System}
\acro{MODIS}{MODerate-resolution Imaging Spectroradiometer}
\acro{MOU}{Memorandum of Understanding}
\acro{MPARWG}{Metrics Planning and Reporting Working Group}
\acro{MSE}{mean square error}
\acro{MSG}{Meteosat Second Generation}
\acro{MTPE}{Mission To Planet Earth}
\acro{MUR}{Multi-sensor Ultra-high Resolution}
\acro{MV}{Motor Vessel}

\acro{NAML}{National Association of Marine Laboratories}
\acro{NAHDO}{National Association of Health Data Organizations}
\acro{NAS}{Network Attached Storage}
\acro{NASA}{National Aeronautics and Space Administration}
\acro{NCAR}{National Center for Atmospheric Research}
\acro{NCEI}{National Centers for Environmental Information}
\acro{NCEP}{National Centers for Environmental Prediction}
\acro{NCDC}{National Climatic Data Center}
\acro{NCDDC}{National Coastal Data Development Center}
\acro{NCL}{NCAR Command Language}
\acro{ncBrowse}{}
\acro{NcML}{\textsmaller{netCDF} Markup Language}
\acro{NCO}{\textsmaller{netCDF} Operator}
\acro{NCODA}{Navy Coupled Ocean Data Assimilation}
\acro{NCSA}{National Center for Supercomputing Applications}
\acro{NDBC}{National Data Buoy Center}
\acro{NDVI}{Normalized Difference Vegetation Index}
\acro{NEC}{North Equatorial Current}
\acro{NECC}{North Equatorial Countercurrent}
\acro{NEFSC}{Northeast Fisheries Science Center}
\acro{NEIC}{North Equatorial Intermediate Current}
\acro{netCDF}{NETwork Common Data Format}
\acro{NEUC}{North Equatorial Undercurrent}
\acro{NGDC}{National Geophysical Data Center}
\acro{NICC}{North Intermediate Countercurrent}
\acro{NIST}{National Institute of Standards and Technology}
\acro{NLSST}{Non-Linear Sea Surface Temperature}
\acro{NMFS}{National Marine Fisheries Service}
\acro{NMS}{New Media Studio}
\acro{NN}{Neural Network}
\acro{NOAA}{National Oceanic and Atmospheric Administration}
\acro{NODC}{National Oceanographic Data Center}
\acro{NOGAPS}{Navy Operational Global Atmospheric Prediction System}
\acro{NOMADS}{\textsmaller{NOAA} Operational Model Archive Distribution System}
\acro{NOPP}{National Oceanographic Parternership Program}
\acro{NOS}{National Ocean Service}
\acro{NPP}{National Polar-orbiting Partnership}
\acro{NPOESS}{National Polar-orbiting Operational Environmental Satellite System}
\acro{NSCAT}{\textsmaller{NASA} SCATterometer}
\acro{NSEN}{\textsmaller{NASA} Science and Engineering Network}
\acro{NSF}{National Science Foundation}
\acro{NSIPP}{NASA Seasonal-to-Interannual Prediction Project}
\acro{NRA}{NASA Research Announcement}
\acro{NRC}{National Research Council}
\acro{NRL}{Naval Research Laboratory}
\acro{NSCC}{North Subsurface Countercurrent}
\acro{NSF}{National Science Foundation}
\acro{NSIDC}{National Snow and Ice Data Center}
\acro{NSPIRES}{\textsmaller{NASA} Solicitation and Proposal Integrated Review and Evaluation System}
\acro{NSSDC}{National Space Science Data Center}
\acro{NVODS}{National Virtual Ocean Data System}
\acro{NWP}{Numerical Weather Prediction}
\acro{NWS}{National Weather Service}

\acro{OBPG}{Ocean Biology Processing Group}
\acro{OB.DAAC}{Ocean Biology \textsmaller{DAAC}}
\acro{ODC}{\textsmaller{OPeNDAP} Data Connector}
\acro{OC}{ocean color}
\acro{OCAPI}{\textsmaller{OPeNDAP C API}}
\acro{ODSIP}{Open Data Services Invocation Protocol}
\acro{OFES}{Ocean Model for the Earth Simulator}
\acro{OCCA}{OCean Comprehensive Atlas}
\acro{OGC}{Open Geospatial Consortium}
\acro{OGCM}{ocean general circulation model}
\acro{OISSTv1}{Optimally Interpolated SST Version 1}
\acro{ONR}{Office of Naval Research}
\acro{OLCI}{Ocean Land Colour Instrument}
\acro{OLFS}{\textsmaller{OPeNDAP} Lightweight Front-end Server}
\acro{OOD}{out-of-distribution}
\acro{OOPC}{Ocean Observation Panel for Climate}
\acro{OPeNDAP}{Open source Project for a Network Data Access Protocol}
\acro{OPeNDAPg}{\textsmaller{GRID}-enabled \textsmaller{OPeNDAP} tools}
\acro{OpenGIS}{OpenGIS}
\acro{OSI SAF}{Ocean and Sea Ice Satellite Application Facility}
\acro{OSS}{Office of Space Science}
\acro{OSTM}{Ocean Surface Topography Mission }
\acro{OSU}{Oregon State University}
\acro{OS X}{}
\acro{OWL}{Web Ontology Language}
\acro{OWASP}{Open Web Application Security Project}

\acro{PAE}{probabilistic autoencoder}
\acro{PBL}{planetary boundary layer}
\acro{PCA}{Principal Components Analysis}
\acro{PDistF}[PDF]{probability distribution function}
\acro{PDF}{probability density function}
\acro{PF}{Polar Front}
\acro{PFEL}{Pacific Fisheries Environmental Laboratory}
\acro{PI}{Principal Investigator}
\acro{PIV}{Particle Image Velocimetry}
\acro{PL}{Project Leader}
\acro{PM}{Project Member}
\acro{PMEL}{Pacific Marine Environmental Laboratory}
\acro{POC}{particulate organic carbon}
\acro{PO-DAAC}{Physical Oceanography -- Distributed Active Archive Center}
\acro{POP}{Parallel Ocean Program }
\acro{PSD}{Power Spectral Density}
\acro{PSPT}{Precision Solar Photometric Telescope}
\acro{PSU}{Pennsylvania State University}
\acro{PyDAP}{Python Data Access Protocol}
\acro{PV}{potential vorticity}

\acro{QC}{quality control}
\acro{QG}{quasi-geostrophic}
\acro{QuikSCAT}{Quick Scatterometer}
\acro{QZJ}{quasi-zonal jet}

\acro{R2HA2}{Rescue \& Reprocessing of Historical AVHRR Archives }
\acro{RAFOS}{{\smaller SOFAR}, SOund Fixing And Ranging, spelled backward}
\acro{RAID}{Redundant Array of Independent Disks}
\acro{RAL}{Rutherford Appleton Laboratory}
\acro{RDF}{Resource Description Language}
\acro{REASoN}{Research, Education and Applications Solutions Network}
\acro{REAP}{Realtime Environment for Analytical Processing}
\acro{ReLU}{Rectified Linear Unit}
\acro{REU}{Research Experiences for Undergraduates}
\acro{RFA}{Research Focus Area}
\acro{RFI}{Radio Frequency Interference}
\acro{RFC}{Request For Comments}
\acro{R/GTS}{Regional/Global Task Sharing}
\acro{RSI}{Research Systems Inc.}
\acro{RISE}{Radiative Inputs from Sun to Earth}
\acro{rms}{root mean square}
\acro{RMI}{Remote Method Invocation}
\acro{ROMS}{Regional Ocean Modeling System}
\acro{ROSES}{Research Opportunities in Space and Earth Sciences}
\acro{RSMAS}{Rosenstiel School of Marine and Atmospheric Science}
\acro{RSS}{Remote Sensing Sytems}
\acro{RTM}{radiative transfer model}

\acro{SACCF}{Southern {\smaller ACC} Front}
\acro{SAC-D}[(SAC)-D]{Sat\'elite de Aplicaciones Cient\'ificas-D}
\acro{SAF}{Subantarctic Front}
\acro{SAIC}{Science Applications International Corporation}
\acro{SANS}{SysAdmin, Audit, Networking, and Security}
\acro{SAR}{synthetic aperature radar}
\acro{SATMOS}{Service d'Archivage et de Traitement M\'et\'eorologique des Observations Spatiales}
\acro{SBE}{Sea-Bird Electronics}
\acro{SciDAC}{Scientific Discovery through Advanced Computing}
\acro{SCC}{Subsurface Countercurrent}
\acro{SCCWRP}{Southern California Coastal Water Research Project}
\acro{SDAC}{Solar Physics Data Analysis Center}
\acro{SDS}{Scientific Data Set}
\acro{SDSC}{San Diego Supercomputer Center}
\acro{SeaDAS}{\textsmaller{SeaWiFS} Data Analysis System}
\acro{SeaWiFS}{Sea-viewing Wide Field-of-view Sensor}
\acro{SEC}{South Equatorial Current}
\acro{SECC}{South Equatorial Countercurrent}
\acro{SECDDS}{Sun Earth Connection Distributed Data Services}
\acro{SEEDS}{Strategic Evolution of \textsmaller{ESE} Data Systems}
\acro{SEIC}{South Equatorial Intermediate Current}
\acro{SEUC}{Southern Equatorial Undercurrent}
\acro{SEVIRI}{Spinning Enhanced Visible and Infra-Red Imager}
\acro{SeRQL}{SeRQL}
\acro{SGI}{Silican Graphics Incorporated}
\acro{SHF}{Sensible Heat Flux}
\acro{SICC}{South Intermediate Countercurrent}
\acro{SIED}{single image edge detection}
\acro{SIPS}{Science Investigator--led Processing System}
\acro{SIR}{Scatterometer Image Reconstruction}
\acro{SISTeR}{Scanning Infrared Sea Surface Temperature Radiometer}
\acro{SLA}{sea level anomaly}
\acro{SMAP}{Soil Moisture Active Passive}
\acro{SMMR}{Scanning Multichannel Microwave Radiometer}
\acro{SMOS}{Soil Moisture and Ocean Salinity}
\acro{SMTP}{Simple Mail Transfer Protocol}
\acro{SOAP}{Simple Object Access Protocol}
\acro{SOEST}{School of Ocean and Earth Science and Technology}
\acro{SOFAR}{SOund Fixing And Ranging}
\acro{SOFINE}{Southern Ocean Finescale Mixing Experiment}
\acro{SOHO}{Solar and Heliospheric Observatory}
\acro{SPARC}{Space Physics and Aeronomy Research Collaboratory}
\acro{SPARQL}{Simple Protocol and \textsmaller{RDF} Query Language}
\acro{SPASE}{Space Physics Archive Search Engine}
\acro{SPCZ}{South Pacific Convergence Zone}  
\acro{SPDF}{Space Physics Data Facility}
\acro{SPDML}{Space Physics Data Markup Language}
\acro{SPG}{Standards Process Group}
\acro{SQL}{Structured Query Language}
\acro{SSCC}{South Subsurface Countercurrent}
\acro{SSL}{Secure Sockets Layer}
\acro{SSO}{Single sign-on}
\acro{SSES}{Single Sensor Error Statistics}
\acro{SSH}{sea surface height}
\acro{SSHA}{sea surface height anomaly}
\acro{SSMI}{Special Sensor Microwave/Imager}
\acro{SST}{sea surface temperature}
\acro{SSTa}{\ac{SST}a}
\acro{SSS}{sea surface salinity}
\acro{SSTST}{Sea Surface Temperature Science Team}
\acro{STEM}{science, technology, engineering and mathematics}
\acro{STL}{Standard Template Library}
\acro{STCZ}{Subtropical Convergence Zone}
\acro{Suomi-NPP}[{\larger Suomi}-NPP]{Suomi-\acl{NPP}}
\acro{SWEET}{Semantic Web for Earth and Environmental Terminology}
\acro{SWFSC}{Southwest Fisheries Science Center}
\acro{SWOT}{Surface Water and Ocean Topography}
\acro{SWRL}{Semantic Web Rule Language}
\acro{SubEx}{Submesoscale Experiment}
\acro{SURA}{Southeastern Universities Research Association}
\acro{SURFO}{Summer Undergraduate Research Fellowship Program in Oceanography}
\acro{SuperDARN}{Super Dual Auroral Radar Network}

\acro{TAMU}{Texas A\&M University}
\acro{TB}{TeraByte - $10^{12}$ bytes}
\acro{TCASCV}{Technology Center for Advanced Scientific Computing and Visualization}
\acro{TCP}{Transmission Control Protocol}
\acro{TCP/IP}{Transmission Control Protocol/Internet Protocol}
\acro{TDS}{\textsmaller{THREDDS} Data Server}
\acro{TEX}{external temperature or T-External}
\acro{THREDDS}{Thematic Realtime Environmental Data Distributed Services}
\acro{TIDI}{\textsmaller{TIMED} Doppler Interferometer}
\acro{TIFF}{Tag Image File Format}
\acro{TIMED}{Thermosphere, Ionosphere, Mesosphere, Energetics and Dynamics}
\acro{TLS}{Transport Layer Security}
\acro{TRL}{Technology Readiness Level}
\acro{TMI}{\textsmaller{TRMM} Microwave Imager}
\acro{TOPEX/Poseidon}{TOPography EXperiment for Ocean Circulation/Poseidon}
\acro{TRMM}{Tropical Rainfall Measuring Mission}
\acro{TSG}{thermosalinograph}

\acro{UCAR}{University Corporation for Atmospheric Research}
\acro{UCSB}{University of California, Santa Barbara}
\acro{UCSD}{University of California, San Diego}
\acro{uCTD}{Underway Conductivity, Temperature and Salinity or Underway \acs{CTD}}
\acro{UDDI}{Universal Description, Discovery and Integration}
\acro{UMAP}{Uniform Manifold Approximation and Projection}
\acro{UMiami}{University of Miami}
\acro{Unidata}{}
\acro{URI}{University of Rhode Island}
\acro{UPC}{Unidata Program Committee}
\acro{URL}{Uniform Resource Locator}
\acro{USGS}{United States Geological Survey}
\acro{UTC}{Coordinated Universal Time}
\acro{UW}{University of Washington}

\acro{VCDAT}{Visual Climate Data Analysis Tools}
\acro{VIIRS}{Visible-Infrared Imager-Radiometer Suite}
\acro{VR}{Virtual Reality}
\acro{VSTO}{Virtual Solar-Terrestrial Observatory}

\acro{WCR}{Warm Core Ring}
\acro{WCS}{Web Coverage Service}
\acro{WCRP}{World Climate Research Program}
\acro{WFS}{Web Feature Service}
\acro{WMS}{Web Map Service}
\acro{W3C}{World Wide Web Consortium}
\acro{WJ}{Wyrtki Jets}
\acro{WHOI}{Woods Hole Oceanographic Institution}
\acro{WKB}{Well Known Binaries}
\acro{WIMP}{Windows, Icons, Menus, and Pointers}
\acro{WIS}{World Meteorological Organisation Information System}
\acro{WOA05}{World Ocean Atlas 2005}
\acro{WOCE}{World Ocean Circulation Experiment}
\acro{WP-ESIP}{Working Prototype Earth Science Information Partner}
\acro{WRF}{Weather \& Research Forecasting Model}
\acro{WSDL}{Web Services Description Language}
\acro{WSP}{western South Pacfic}
\acro{WWW}{World Wide Web}

\acro{XBT}{Expendable BathyThermograph}
\acro{XML}{Extensible Markup Language}
\acro{XRAC}{eXtreme Digital Request Allocation Committee}
\acro{XSEDE}{Extreme Science and Engineering Discovery Environment}

\acro{YAG}{yttrium aluminium garnet}

\end{acronym}

\end{doublespace}




\newpage

\centerline{CONTRIBUTIONS}
Prochaska led the writing of the manuscript, including figure generation.
He also ran the majority of models presented. Cornillon proposed the original idea of searching for extremes in the \ac{MODIS} \ac{L2} {\sst} dataset, undertook a significant fraction of the analysis of the resulting \ac{LL} fields, guided the oceanographic interpretation of the results and contributed to the writing of the manuscript. Reiman prototyped and developed \ulmo's deep learning components.

\bigskip
\centerline{ACKNOWLEDGMENTS}
JXP recognizes support from the University of California, Santa Cruz.
Support for PC was provided by the Office of Naval Research: \acs{ONR} N00014-17-1-2963 and \ac{NASA} grant \# 80NSSC18K0837
MODIS \ac{SST} data were produced by and obtained from the NASA Goddard Space Flight Center, Ocean Ecology Laboratory, Ocean Biology Processing Group; 
bathymetry was obtained from the GEBCO Compilation Group (2020) GEBCO 2020 Grid (doi:10.5285/a29c5465-b138-234d-e053-6c86abc040b9); the authors gratefully acknowledge helpful discussions with Baylor Fox-Kemper of Brown University, Chris Edwards of UC Santa Cruz and Peter Minnett and Kay Kilpatrick of the University of Miami.
\bigskip



\bibliography{bibliography}



\end{document}